\documentclass[12pt]{article}

\usepackage[dvips]{graphicx}
\usepackage{epsfig}
\usepackage{amsmath,amsfonts,amssymb,amsthm}
\usepackage{mathrsfs,mathtools}
\usepackage{verbatim}
\usepackage{psfrag}
\usepackage{bm}
\usepackage{bbm}
\usepackage[utf8]{inputenc}
\usepackage[square,comma,sort&compress,numbers]{natbib}
\usepackage[dvipsnames]{xcolor}
\usepackage{slashed}
\usepackage{upgreek}
\usepackage{enumitem}
\usepackage{float}
\usepackage[T1]{fontenc}
\usepackage{soul}
\usepackage{pgfplots}
\usepackage{extarrows}
\pgfplotsset{compat=1.18}
\usetikzlibrary{external,decorations.pathmorphing}
\usepackage{cases}
\usepackage{empheq}
\usepackage{physics}
\usepackage{epsf,epsfig}
\usepackage{graphics}
\usepackage{subcaption}
\usepackage{lmodern}
\usepackage{tikz}
\usepackage[normalem]{ulem}
\usepackage[colorlinks=true,urlcolor=blue,anchorcolor=blue,citecolor=blue,filecolor=blue
,linkcolor=blue,menucolor=blue,linktocpage=true,pdfproducer=medialab,pdfa=true
]{hyperref}
\usepackage[T1]{fontenc}
\usepackage[errorstop]{feynmp}
\usepackage{lettrine}
\usepackage{Zallman}

\tikzset{
  external/only named=true,
  thick/.style={line width=.5pt},
  approximation/.style={line width=1.2pt},
  numerics/.style={black, dotted, line width=.8pt},
  amplitude/.style={dashed},
  estimate/.style={dashed, line width=.8pt},
  normal plot/.style={line width=.8pt},
}

\tikzset{snake it/.style={decorate, decoration=snake}}

\newcommand\blfootnote[1]{%
  \begingroup
  \renewcommand\thefootnote{}\footnote{#1}%
  \addtocounter{footnote}{-1}%
  \endgroup
}

\def\d{\mathrm{d}}
\def\D{\mathcal{D}}
\def\A{\mathcal{A}}
\def\L{\mathcal{L}}
\def\H{\mathcal{H}}
\def\C{\mathcal{C}}
\def\O{\mathcal{O}}
\def\vec{\mathbf}
\def\i{\mathrm{i}}
\def\e{\mathrm{e}}

\def\sign{\rm sign}

\def\M{\mathcal{M}}
\def\veck{\vec{k}}

\def\vecx{\vec{x}}

\newcommand{\gsim}
{\;\raisebox{-.3em}{$\stackrel{\displaystyle >}{\sim}$}\;}

\newcommand{\rom}[1]{\uppercase\expandafter{\romannumeral #1\relax}}
\newcommand{\overbar}[1]{\mkern 1.5mu\overline{\mkern-1.5mu#1\mkern-1.5mu}\mkern 1.5mu}

\begin{document}                                                                                                                      

\thispagestyle{empty}

\begin{flushright}
{
\small
KCL-PH-TH/2023-53\\
}
\end{flushright}

\vspace{-0.5cm}

\begin{center}
\Large\bf\boldmath
From QFT to Boltzmann: Freeze-in in the presence of oscillating condensates
\unboldmath
\end{center}

\vspace{-0.2cm}

\begin{center}
Wen-Yuan Ai,$^*$\blfootnote{$^*$wenyuan.ai@kcl.ac.uk (corresponding author)}
Ankit Beniwal,$^\dagger$\blfootnote{$^\dagger$ankit.beniwal2@gmail.com}
Angelo Maggi$^\ddagger$\blfootnote{$\dagger$angelo.maggi@kcl.ac.uk} 
and David J. E. Marsh$^\S$\blfootnote{$^\S$david.j.marsh@kcl.ac.uk}
\vskip0.4cm

{\it Theoretical Particle Physics and Cosmology, King’s College London,\\ Strand, London WC2R 2LS, UK}
\vskip1.cm
\end{center}

\begin{abstract}
Scalar dark matter (DM), and axions in particular, have an irreducible abundance of particles produced by freeze-in due to portal interactions with the Standard Model plasma in the early Universe. In addition, vacuum misalignment and other mechanisms can lead to the presence of a cold, oscillating condensate. Therefore, generically, the evolution of the DM in both forms, condensate and particles, needs to be studied simultaneously. In non-equilibrium quantum field theory, the condensate and particles are described by one- and two-point functions, respectively. The fundamental coupled equations of motion (EoMs) of these objects are non-local. To simplify the EoMs and bring them into a familiar form for relic abundance calculations, we perform a Markovianization process for a quasi-harmonically oscillating homogeneous condensate, leading to local EoMs for the particle distribution function and the envelope function of condensate oscillation. This reduces the dynamics to a pair of coupled Boltzmann equations, and we derive explicitly the form of the collision operators for all particle and condensate interactions.
\end{abstract}

\newpage
%
\hrule
\tableofcontents
\vskip.85cm
\hrule


\section{Introduction and summary of the results}
\label{sec:Intro}

\lettrine{D}ark Matter (DM) is known to compose around 85\% of the matter content of the Universe and is necessary for the large-scale structure of the Universe and the formation of galactic halos~\cite{Planck:2018vyg}. DM can range in mass from the lightest scalar fields with $m\approx 10^{-22}\,\text{eV}$, to the heaviest primordial black holes with mass $M\approx 10^4 M_\odot$ (for an overview, see Ref.~\cite{AlvesBatista:2021eeu}). Axions~\cite{Peccei:1977hh,Wilczek:1977pj,Weinberg:1977ma,Abbott:1982af,Preskill:1982cy,Dine:1982ah}, and axion-like particles~\cite{Arias:2012az,Marsh:2015xka} (henceforth, axions) generally occupy the lower end of this mass spectrum of candidates and motivate the present work.

Scalar fields, $\Phi$, like the axion, can be produced in the early Universe via ``vacuum misalignment'' with no need for non-gravitational interactions with the Standard Model (SM)~\cite{Abbott:1982af,Preskill:1982cy,Dine:1982ah,Turner:1983he}. In this process, the scalar condensate, $\varphi$, is displaced from the minimum of its potential. This displacement may be due to spontaneous symmetry breaking, or fluctuations during inflation~\cite{Linde:1982uu,Affleck:1984fy,Starobinsky:1994bd}. Hubble friction keeps the condensate held at the displaced value until $H\lesssim M_\phi$  (where $H$ is the Hubble parameter and $M_\phi$ is the thermally corrected scalar mass) when the condensate begins to oscillate around the minimum. Energy density in the condensate scales like $\rho_\varphi\sim a^{-3}$ (where $a$ is the cosmic scale factor), thus behaving as cold DM. This mechanism has two free parameters, the initial displacement, $\varphi_i$, and the scalar mass, $M_\phi$, and thus can provide the correct relic density for any mass of the scalar field.

On the other hand, scalar particles can be produced by their interactions with the thermal SM plasma. In the case of axions, the main interaction is the \emph{Primakoff interaction}: scattering of photons off charged particles, creating a single axion. Even if the maximum temperature is so low that such an interaction cannot thermalize, an irreducible contribution is created by the freeze-in process~\cite{Hall:2009bx,Balazs:2022tjl,Langhoff:2022bij}. In the following we consider a general case where freeze-in is mediated by some portal field, $\chi$ in thermal equilibrium with the SM. Once the resulting equations are derived, generalisation to other portal interactions is possible by writing down the relevant Feynman diagrams.

It is commonplace in the DM literature to treat the DM particles and condensate separately, possibly neglecting one or the other, or simply treating the combined relic abundance as a free parameter. In the following, we explore in detail the interaction between condensates and particles. We shall use the closed-time-path (CTP) formalism~\cite{Schwinger:1960qe,Keldysh:1964ud} (also called Schwinger-Keldysh or in-in formalism) to derive the coupled equations. The derivation of Boltzmann equations for particles is well known in the literature on non-equilibrium quantum field theory (QFT) (see e.g., Refs.~\cite{Calzetta:1986cq,Prokopec:2003pj,Prokopec:2004ic,Drewes:2012qw,Sheng:2021kfc}). However, the situation when condensates are present is less explored. Furthermore, the formalism can obscure the essential physics for non-experts working in the field of DM phenomenology. We thus aim at a pedagogical derivation of particle-condensate interactions, with an end product that is easy to use and generalize in relic density calculations. We anticipate that our results will be particularly applicable to models with multiple axions, where the abundances of particles and condensates can be comparable, and where interactions between multiple species are present \cite{Gendler:2023kjt}.

The condensate is described by the one-point function, $\varphi\equiv \langle \Phi\rangle$, where the brackets denote the expectation value. Being an expectation value, $\varphi$ behaves much like a classical field but its evolution receives quantum and thermal corrections. The latter are important as they give rise to the dissipation of the condensate. Since the condensate carries an energy density $\rho_\varphi$, $\rho_\varphi/M_\phi$ can be interpreted as a {\it number density of condensate quanta}, $n_\varphi$. 

\begin{figure}[t]
    \centering
    \includegraphics[scale=0.6]{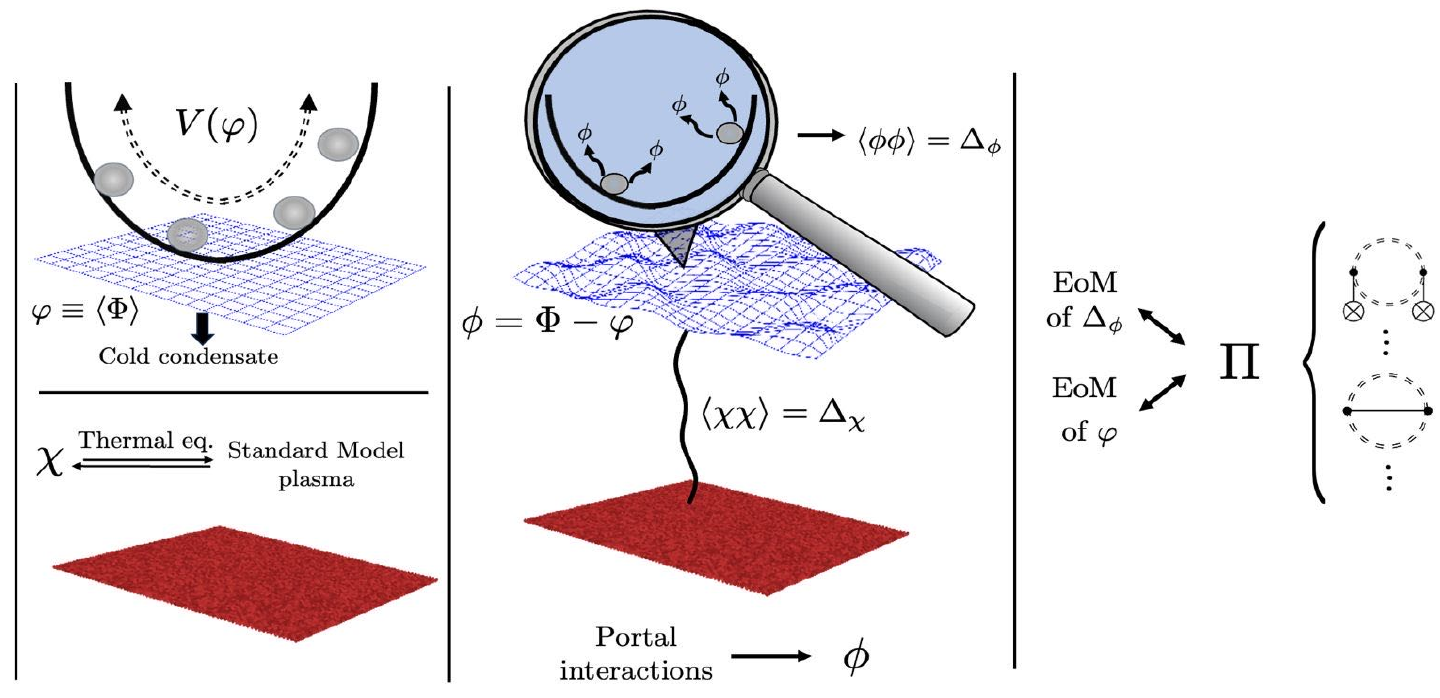}
    \caption{This cartoon illustrates the phenomenon we study in this paper: the evolution of an oscillating homogeneous condensate $\varphi$ of a DM (quantum) field $\Phi$ (hence $\varphi= \langle \Phi\rangle$) in a SM thermal plasma. The condensate interacts with the plasma through a portal scalar $\chi$ which is in thermal equilibrium with the plasma. The fluctuation field upon the condensate $\phi=\Phi-\varphi$ describes DM particles. Generically, particles are described by the connected two-point functions $\Delta_{\phi,\,\chi}$. The equations of motion (EoMs) for the condensate $\varphi$ and the two-point function $\Delta_\phi$ ($\Delta_\chi$ is assumed to take the thermal equilibrium form) involve various self-energies $\Pi$ that characterize the interactions and can be represented by Feynman diagrams. The fundamental equations are non-local, essentially due to $\Delta$ and $\Pi$ carrying two coordinates.}
    \label{fig:cartoon}
\end{figure}

In the background of the condensate, particle excitations are described by the {\it fluctuation} field $\phi=\Phi-\langle \Phi\rangle $. Indeed, $\phi$ inherits the operator character of the field $\Phi$ and thus behaves as a quantum field. Since $\langle\phi\rangle=0$, the information of particles is encoded in the two-point function of $\phi$, namely $\Delta_\phi\equiv \langle \phi\phi\rangle=\langle\Phi\Phi\rangle-\varphi\varphi=\langle \Phi\Phi\rangle_c$ where the subscript ``c'' stands for ``connected''. For example, the energy-momentum tensor of the $\phi$-particles can be obtained, at the leading order, from the two-point function $\Delta_\phi$.\footnote{In principle, one can have an infinite set of coupled EoMs for all correlation functions. This is the relativistic quantum field theoretical generalization of Bogoliubov-Born-Green-Kirkwood-Yvon hierarchy. Usually, it is sufficient to truncate this series up to the EoMs for the two-point functions and this leads to  dissipative characteristics in the subsystem. Higher $n$-point correlation functions are then calculated as perturbative series in terms of two-point functions.} Thus, the latter must be related to the Boltzmann distribution function, $f_{\phi}(k,x)$, which gives the classical probability for a particle to occupy a region in phase space ${\rm d}^3 x{\rm \, d}^3k$.  The two-particle irreducible (2PI) effective action~\cite{Cornwall:1974vz} in QFT can be used to derive coupled equations of motion (EoMs) for the one- and two-point functions, $\varphi$ and $\Delta_\phi$. The EoMs contain integrals over the entire past lightcone involving various self-energies, broadly denoted as $\Pi$ here, and thus are \emph{non-local}. For oscillating but homogeneous condensate backgrounds, the EoMs are thus non-Markovian (non-local in time). Fig.~\ref{fig:cartoon} illustrates the physical picture for particle and condensate dynamics.

The classical non-relativistic Boltzmann equations that are typically used to understand DM production in the early Universe are Markovian, which makes them easily tractable to find the relic density. The details of Markovianization, and the relationship between non-equilibrium QFT and the Boltzmann equations are derived in detail in the following, but we briefly summarise them here. 

For the EoM of $\Delta_\phi$, one first writes a general non-local two-point correlation function, $\Sigma(x_1,x_2)$, which can be $\Delta_\phi$ or self-energies, as $\Sigma(x=x_1/2+x_2/2, \,r=x_1-x_2)$. The Fourier transform with respect to $r$, which is called the Wigner transform, gives $\overbar{\Sigma}(k,x)$. In Wigner space $\{x,k\}$, $k$ can be viewed as (microscopic) particle momentum at $x$, where $x$ characterizes the bulk motion. The convolution of two two-point functions is transformed into a gradient expansion in Wigner space. Truncating at a finite order, one then arrives at a localized EoM.

To obtain a Boltzmann equation for the particle distribution, the appropriate two-point function to consider is the \emph{Wigner transform of the Wightman function}, $\overbar{\Delta}_{\phi}^<(k,x)$ (or $\overbar{\Delta}_{\phi}^>(k,x)$). In the on-shell limit, where the quasi-particle picture can be used, the Wightman function is given by~\cite{Prokopec:2003pj}:
\begin{equation}
\overbar{\Delta}_{\phi}^<(k,x)=2\pi \delta (k^2-M^2_{\phi})\,{\sign}(k_0)f_{\phi}(k,x)\,,
\end{equation}
where $M_\phi$ contains quantum and thermal corrections.

For an oscillating condensate, the aforementioned gradient expansion is usually not sufficient. In this case, one can use multiple-scale analysis~\cite{Ai:2021gtg,Wang:2022mvv,Ai:2023ahr} to localize the EoM of a quasi-harmonically oscillating condensate leading to a solution in the form
\begin{equation}
    \varphi(t) \approx A(t)\cos (M_\phi t) \rightarrow n_\varphi \approx\frac{1}{2}M_\phi A^2\,.
\end{equation}
One then reduces the non-local condensate EoM to a local one for the unknown envelop function $A(t)$~\cite{Ai:2023ahr}, or equivalently $n_\varphi(t)$.

\paragraph{Summary of the results.}

We assume a homogeneous and isotropic universe such that the background spacetime is described by the Friedmann-Lemaître-Robertson-Walker (FLRW) metric. After a long derivation, we end up with a set of simple coupled Boltzmann equations
\begin{subequations}
\label{eq:results}
\begin{align}
\label{eq:Boltzman-condensate}
&\dot{n}_\varphi+3H n_\varphi=\hat{\C}_\pi[f_i] n_\varphi +2  \hat{\C}_v[f_i] n_\varphi^2\,,\\
\label{eq:Boltz-number-density}
&\dot{n}_{\phi} + 3 H n_{\phi}(t) =\int \frac{\d^3 \veck}{(2\pi)^3}\C[f_{\phi}(\veck,t),f_{i};n_\varphi]\,,
\end{align}
\end{subequations}
where $n_\phi\equiv\int\frac{\d^3\veck}{(2\pi)^3} f_\phi(\veck,t)$ is the {\it particle} number density for the dark matter, and $f_i$ generally denotes the distribution function for any particle species that are involved in the interaction. We also recall that $n_\varphi=\rho_\varphi/M_\phi$ is the number density of condensate quanta. The total DM density is then given by $M_\phi(n_\phi+n_\varphi)$. $\hat{\C}_{\pi/v}$ are collision terms that describe the interaction involving condensate quanta and particles. The power on $n_\varphi$ depends on the power of $\varphi$ in the vertex considered in the collision operators $\hat{\C}_{\pi/v}$, and physically corresponds to how many condensate quanta are involved in the interaction (cf. Eq.~\eqref{eq:collision-term-condensate}). The collision operator $\C$ can be classified into two classes: $(i)$ The vertex considered contains only fluctuation fields; in this case $\C$ is simply the standard familiar collision operator for interactions among particles. $(ii)$ The vertex contains the condensate field $\varphi$. In this case, in the interaction there are condensate quanta emitted or absorbed and this type of $\C$ is discussed specifically in Sec.~\ref{sec:collision-with-condensate}.

The outcome of our derivation, namely that both the condensate and particles can be described by Boltzmann equations, is reminiscent of some works in condensed matter physics, e.g., Refs.~\cite{nikuni1999two,zaremba1999dynamics}. These works are based on local (non-relativistic) EoM for the condensate. The locality there is essentially due to the Hartree approximation used. Going beyond that approximation, one has to deal with non-local EoMs as we do in this paper. Our derivation might be viewed as a relativistic and more complete derivation of the coupled dynamics between condensates and particles. The dynamics can be applied, though not commonly, to the interaction between the condensate and particles during the formation of Bose-Einstein condensates at finite temperature in condensed matter systems (e.g., Ref.~\cite{bao2013mathematical}).  We also note that recently a different description on the condensate-particle system is given in Refs.~\cite{Proukakis:2023nmm,Proukakis:2023txk}. Although the authors in these references also use the CTP formalism, they identify the ``condensate'' and ``particle'' differently such that both condensate and particle fields have classical and quantum parts. 

The remainder of the paper is to provide all the details of the derivation of Eqs~\eqref{eq:results} from first principles, starting from a simple two-scalar model. The derivation itself, however, is quite general and should be easily generalised to more complicated models. We do not apply the derived equations to phenomenology in this paper but satisfy ourselves by generalising the discussion to a more realistic axion model, focusing on the important Primakoff interaction. This paper is organized as follows: in section~\ref{sec:model}, we introduce the model under study; in section~\ref{sec:CPT-formalism}, we develop the CTP formalism; in section~\ref{sec:nonlocal-EoM}, we derive the non-local EoMs; in section~\ref{sec:markovianize}, we show how these equations can be Markovianized; in section~\ref{sec:Boltzmann}, the Boltzmann equations are derived finally and generalisation to axion models is discussed; conclusions are given in section~\ref{sec:Conc}. Two appendices give further details on the Feynman rules and cutting rules used in the derivation of the EoMs.

\section{Overview of a simplified model}\label{sec:model}

We consider the following model with a real DM scalar field $\Phi$ and a portal scalar field $\chi$, 
\begin{align}\label{eq:model}
    \L/\sqrt{-g} &=
	\frac{1}{2} (\partial_\mu \Phi) (\partial^\mu \Phi)
	+ \frac{1}{2} (\partial_\mu \chi) (\partial^\mu \chi)
	-V(\Phi,\chi)\,,
\end{align}
where $g$ is the metric determinant with signature $(+,-,-,-)$. We take the interaction potential to be:
\begin{equation}\label{eq:pot}
V(\Phi,\chi) = \frac{1}{2} m_\phi^2 \Phi^2
	+ \frac{1}{2} m_\chi^2 \chi^2
	+ \frac{1}{4!} \lambda_\phi \Phi^4
	+ \frac{1}{4!} \lambda_\chi \chi^4
	+ \frac{1}{4} g \Phi^2 \chi^2\,. \\[1mm]
\end{equation}
The potential allows scattering between particles but does not allow particle decays due to the $\mathbb{Z}_2$ symmetries. This reduces the number of terms that we must keep track of explicitly. We discuss the Primakoff interaction for axions at the end of this paper.

We assume $m_\phi^2>m_\chi^2>0$ and that the DM field $\Phi$ forms a condensate,  i.e., its one-point function $\langle \Phi \rangle \equiv \varphi$ is non-vanishing. In the presence of a condensate background field, DM particles are then defined as excitations of the corresponding fluctuation field upon the background field, $\phi=\Phi-\varphi $. Aside from the fluctuation field $\phi$, we also have the $\chi$ and SM fields, which can also be called fluctuation fields due to their vanishing one-point functions. 

We are interested in processes in the early Universe, which is described by the FLRW metric with scale factor $a(t)$ and Hubble rate, $H=\dot{a}/a$, given by: 
\begin{equation}
    3H^2 M_{\rm P}^2 = \frac{\pi^2}{30}g_\star T^4\, ,
\end{equation}
where $M_{\rm P}$ is the {\it reduced} Planck mass, $T$ is the temperature of the photons, and $g_\star$ is the number of relativistic degrees of freedom in the energy density. As QFT in curved space-time contains many subtleties, we assume a flat space-time at the level of QFT. Only after obtaining the EoMs for the background field $\varphi$ or distribution functions will we consider modifications due to the expansion of the Universe. 

We assume that the coupling between the DM and portal fields is feeble such that DM particles are never in thermal equilibrium, leading to the freeze-in scenario of DM production. The portal field $\chi$ is assumed to be in thermal equilibrium with the SM. This can describe Higgs portal DM (in that case $m_\chi^2<0$ which could bring more complications due to the possible spontaneous symmetry breaking of the $\mathbb{Z}_2$ symmetry of the $\chi$ field) or is intended to be a toy model for the case of axions coupled to the SM by the Primakoff interaction such that $\chi$ is a stand-in for electrons or other charged particles. We explicitly give the Primakoff case in section~\ref{sec:primakoff}.

The presence of the condensate $\varphi$ introduces new interactions, e.g., the condensate decay processes\footnote{In contrast to ordinary particles in vacuum, condensates can decay even when there is a $\mathbb{Z}_2$ symmetry for the scalar field.}, and scattering processes between the condensate and ordinary particles. While the condensate is described by a one-point function, the information about particles is encoded in the connected two-point functions. As we will shortly derive, the EoMs for the one- and two-point functions are coupled integro-differential equations, where the integrals are over the whole past lightcone. The integrals give the equations \emph{memory} thus making them \emph{non-local and non-Markovian}. The general situation is complicated if we allow for an inhomogeneous background field configuration, leading to partial differential equations in $t$ and $\bm{x}$. For the fields that are in thermal equilibrium, i.e., $\chi$ and other SM fields, we can take their two-point functions to be approximated by the corresponding free thermal equilibrium ones, with thermal masses taken into account if necessary. Thus, we will be left with only two coupled EoMs for $\varphi$ and $\Delta_\phi=\langle\Phi\Phi\rangle_c=\langle \phi\phi\rangle$. 

Various simplifications can be made. For the case that $\phi$-particles are also in thermal equilibrium, we only have one EoM for $\varphi$, which describes the dissipation of the condensate; this case has been studied recently in detail in Refs.~\cite{Ai:2021gtg,Wang:2022mvv,Ai:2023ahr} for the same model\footnote{Earlier studies on the dissipation of scalar backgrounds can be found in Refs.~\cite{Calzetta:1989vs,Paz:1990sd,Boyanovsky:1994me,Greiner:1996dx,Yokoyama:2004pf,Bastero-Gil:2010dgy,Bastero-Gil:2012akf,Mukaida:2012qn}. Different from these references, Refs.~\cite{Ai:2021gtg,Wang:2022mvv,Ai:2023ahr} adopt multiple-scale analysis~\cite{Bender,Holmes} to obtain self-consistent solutions for the condensate evolution in the quasi-harmonic oscillation regime.}, and in Refs~\cite{Cao:2022bua,Cao:2022kjn} for axion condensates.  In the following, we allow for $\phi$ particles to be out of equilibrium, motivated by the freeze-in production scenario for axion DM. We simplify the system by assuming homogeneity and show how to Markovianize the EoMs to bring them into the form of Boltzmann equations (see e.g., Refs.~\cite{Calzetta:1986cq,Ivanov:1999tj,Buchmuller:2000nd,Blaizot:2001nr,Prokopec:2003pj,Prokopec:2004ic,Berges:2005md,DeSimone:2007gkc,Cirigliano:2009yt,Beneke:2010wd,Beneke:2010dz,Drewes:2012qw,Berges:2014xea} for previous work on this).

\begin{figure}[t]
    \centering
    \includegraphics[scale=0.65]{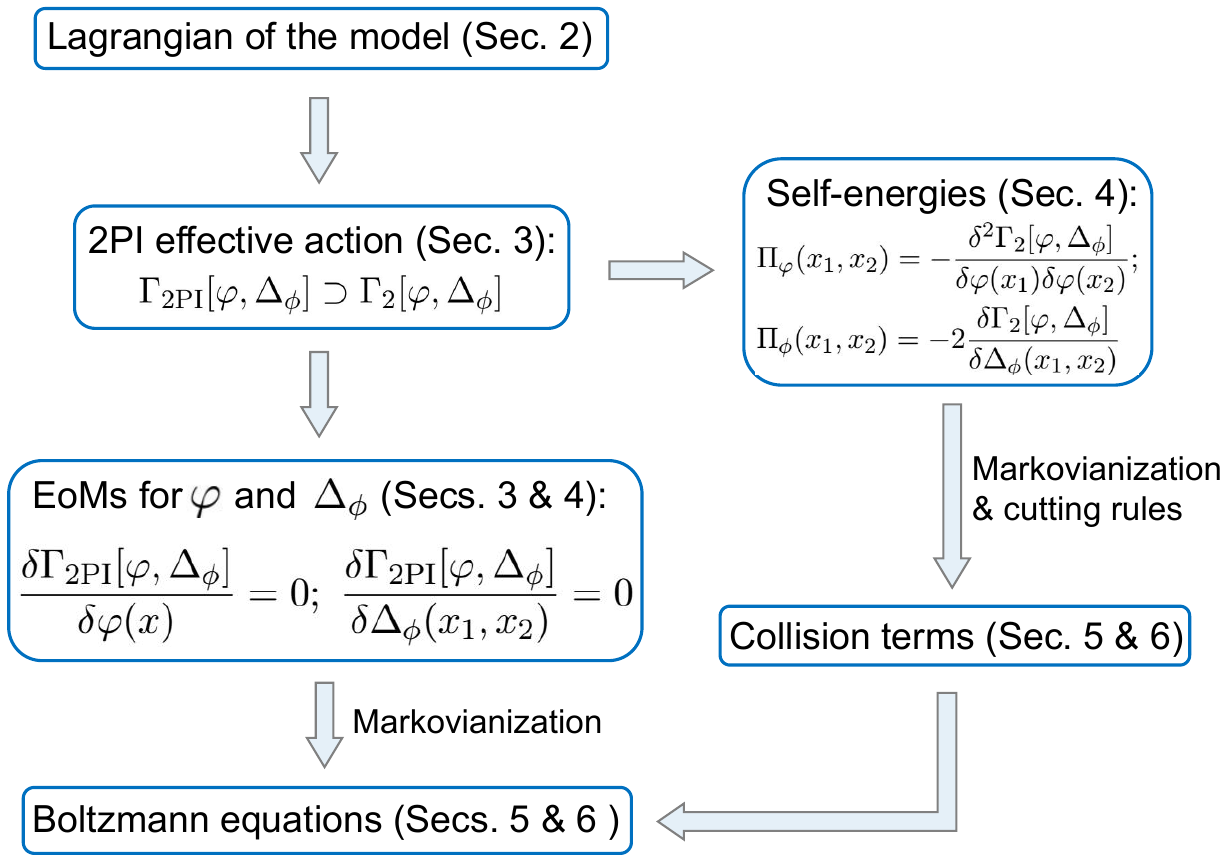}
    \caption{Flowchart of the derivation of Boltzmann equations in the CTP formalism. Here $\Gamma_2$ represents the contribution to the 2PI effective action $\Gamma_{\rm 2PI}$ starting from two loops, and can be represented by two-particle-irreducible Feynman diagrams. We have suppressed the $\pm$ indices from the CTP formalism.}
    \label{fig:flow}
\end{figure}

The CTP formalism that we are going to use is a rather useful theoretical framework where one can study, in principle, arbitrary non-equilibrium phenomena with quantum fields. It has been widely applied to baryogenesis (see, e.g., Refs.~\cite{Buchmuller:2000nd,Prokopec:2003pj,Prokopec:2004ic,Konstandin:2004gy,Lee:2004we,Konstandin:2005cd,DeSimone:2007gkc,Garny:2009rv,Garny:2009qn,Anisimov:2010aq,Beneke:2010wd,Beneke:2010dz,Anisimov:2010dk,BhupalDev:2014oar,Garbrecht:2018mrp,Postma:2022dbr}). Recently, it has also been used to study DM freeze-in (in the absence of a condensate)~\cite{Becker:2023vwd}. For this reason, we perform our derivation in a pedagogical manner and start with a brief introduction to the CTP formalism and 2PI effective action. The technical details may obscure the essential physics. Thus, in Fig.~\ref{fig:flow}, we provide a flowchart to summarise the essence of  the formalism and the structure of the derivation carried out in the remaining sections.

\section{CTP formalism and 2PI effective action}\label{sec:CPT-formalism}

To avoid being distracted by the complications of the CTP formalism,
we shall keep the introduction very brief and refer the interested reader to reviews~\cite{Chou:1984es,Calzetta:1986cq,Berges:2004yj} for more details. The CTP formalism is built on $n$-particle-irreducible effective actions~\cite{Jackiw:1974cv,Cornwall:1974vz} which provide a systematic way to derive {\it quantum} EoMs for correlation functions that can be used both for the conventional zero-temperature quantum field theory and non-equilibrium quantum field theory.\footnote{When the one-point function of the scalar $\Phi$ plays the role of the order parameter in a first-order phase transition, the 1PI or 2PI effective actions can also be used to study radiative corrections to bubble nucleation~\cite{Baacke:1993ne,Surig:1997ne,Baacke:2003uw,Bergner:2003id,Baacke:2006kv,Garbrecht:2015oea,Garbrecht:2015yza,Ai:2018guc,Ai:2020sru,Ai:2023yce}.} Often, 1PI and 2PI effective actions are used. The difference between the 1PI and 2PI effective actions reside in the use of propagators in the perturbative expansion. The 1PI effective action uses free propagators while in the 2PI effective action, the propagators are the full two-point functions that are determined from the EoM.  

We give a brief introduction to the 2PI effective action at zero temperature, followed by a discussion on the complications introduced when going beyond the zero-temperature scenario.

\subsection{\texorpdfstring{2PI}{TEXT} effective action}

Here the 2PI refers to the Feynman diagrams representing the effective action (except for the classical action and one-loop functional determinants) that cannot be separated into two disconnected parts by cutting two propagators, as illustrated in Appendix~\ref{app:Feynman} where we also introduce the Feynman rules for our model in Eq.~\eqref{eq:model}. We illustrate the 2PI effective action using a simpler model with only one scalar field $\Phi$ whose action is denoted as $S[\Phi]$. While the 1PI effective action is a functional of the one-point function $\varphi$, the 2PI effective action is a functional of both $\varphi$ and the two-point function $\Delta_\phi$.

Introducing both local and non-local sources, the generating functional reads
\begin{align}
    Z[J,K]=\int\D\Phi\, \exp \left\{\i\left(S[\Phi]+\int\d^4 x\, J(x)\Phi(x)+\frac{1}{2}\int\d^4 x_1 \, \d^4 x_2\, K(x_1,x_2)\Phi(x_1)\Phi(x_2)\right) \right\}\,.
\end{align}
Consequently,
\begin{subequations}
\begin{align}
  &\varphi(x)=\langle\Phi(x)\rangle= \frac{\delta\ln Z[J,K]}{\i\delta J(x)}\,,\\
  &\Delta_\phi(x_1,x_2)= \langle T\Phi(x_1)\Phi(x_2)\rangle_c=\frac{\delta \ln Z[J,K]}{\i\delta K(x_1,x_2)}-\varphi(x_1)\varphi(x_2)\,.
\end{align}
\end{subequations}
The 2PI effective action is defined as the Legendre transform of $-\i\ln Z[J,K]$,
\begin{align}
    \Gamma_{\rm 2PI}[\varphi,\Delta_\phi] &= -\i\ln Z[J,K]-\int\d^4 x\, J(x)\varphi(x) \nonumber \\
    &\hspace{5mm} -\frac{1}{2}\iint\d^4 x_1 \, \d^4 x_2\, K(x_1,x_2)\left[\Delta_\phi(x_1,x_2)+\varphi(x_1)\varphi(x_2)\right]\,. 
\end{align}
It can be easily checked that
\begin{subequations}
\begin{align}
    &\frac{\delta \Gamma_{\rm 2PI}[\varphi,\Delta_\phi]}{\delta\varphi(x)}=-J(x)-\int\d^4 y\, K(x,y)\varphi(y)\,,\\
    &\frac{\delta \Gamma_{\rm 2PI}[\varphi,\Delta_\phi]}{\delta\Delta_\phi(x_1,x_2)}=-\frac{1}{2} K(x_1,x_2)\,.
\end{align}
\end{subequations}
When there are no external sources, one has 
\begin{align}
    &\frac{\delta \Gamma_{\rm 2PI}[\varphi,\Delta_\phi]}{\delta\varphi(x)}=0\,,\qquad \frac{\delta \Gamma_{\rm 2PI}[\varphi,\Delta_\phi]}{\delta\Delta_\phi(x_1,x_2)}=0\,.
\end{align}
Thus, the 2PI effective action provides the quantum EoM for the one- and two-point functions simultaneously. Meanwhile, the two-point function $\Delta_\phi$ is also used as the propagator in the Feynman-diagram expansion of the 2PI effective action. In addition (see e.g., Ref.~\cite{Berges:2004yj}),
\begin{align}
    \Gamma_{\rm 2PI}[\varphi,\Delta_\phi]=S[\varphi]+\frac{\i}{2}{\rm Tr}\ln \Delta_\phi^{-1}+\frac{\i}{2}{\rm Tr} \left(G_\phi^{-1}\Delta_\phi\right)+\Gamma_2[\varphi,\Delta_\phi]\,,
\end{align}
where $G_\phi^{-1}$ is defined as
\begin{align}
\label{eq:kinetic-operator}
    -\i\left.\frac{\delta^2 S[\Phi]}{\delta\Phi(x_1)\delta\Phi(x_2)}\right|_{\varphi}\equiv \delta^{(4)}(x_1-x_2) G^{-1}_\phi\,,
\end{align}
and $\Gamma_2$ is the contribution starting from the two-loop order:
\begin{align}
   \Gamma_2[\varphi,\Delta_\phi]= -\i\times {\rm the\ sum\ of\ 2PI\ vacuum\ diagrams}\,.
\end{align}
The trace is taken over position space. If the field carries other indices, one also needs to trace over the corresponding inner space, e.g., the spinor space for fermions. 

\subsection{Closed time path}

Now we move onto non-equilibrium QFT. Different from the conventional zero-temperature QFT, where there are known in and out states in the asymptotic past and future, in non-equilibrium QFT, the initial states are prepared and the final states are unknown {\it a priori} and can only be determined by the evolution itself. Thus non-equilibrium dynamics is an initial-value problem. This is the reason why a closed time path needs to be introduced in the generating functional. 

\begin{figure}[t]
    \centering
    \includegraphics[scale=0.3]{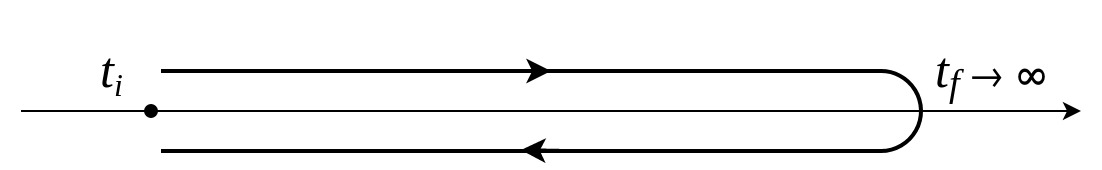}
    \caption{The Keldysh contour $\mathcal{C}$ for the generating functional in the CTP formalism. Here the forward and backward time contours are slightly shifted off the real line only for the purpose of illustration; both contours should be understood as lying exactly on the real line.}
    \label{fig:keldyshcontour}
\end{figure}

The initial value is usually given by a density matrix at a given time $\rho_D(t_i)$ in a mixed (${\rm Tr}\{\rho^2_D(t_i)\}<1$) or pure (${\rm Tr}\{\rho^2_D(t_i)\}=1$) state. In the Heisenberg picture, operators evolve with time while the states do not. The expectation value of a simple observable $\mathcal{O}$ at time $t$ is given by
\begin{align}
\label{eq:expetation}
    \langle \mathcal{O}(t)\rangle ={\rm Tr} \left\{\rho_D(t_i)\mathcal{O}(t)\right\}\,,
\end{align}
where $\mathcal{O}(t)=\exp(\i H(t-t_i))\mathcal{O}(t_i)\exp(-\i H(t-t_i))$. These expectation values can be obtained by a generating functional formulated on a closed time contour $\mathcal{C}$, as illustrated in Fig.~\ref{fig:keldyshcontour}. For the quantity given in Eq.~\eqref{eq:expetation}, $t_f=t$, but one can also choose any $t_f\geq t$ by inserting an operator $\exp(\i H(t-t_f))\exp(-\i H (t_f-t))=\mathbf{1}$ on the left of $\O(t)$ in Eq.~\eqref{eq:expetation}. In particular, one can choose $t_f\rightarrow\infty$ such that the generating functional is independent of the quantities to be computed. This discussion can be generalized to the case of composite operators with multiple time variables $\O(t_1,...,t_n)$.

All the analysis given in the last subsection applies except that the integrals are now performed on a closed time contour. For example, we now have 
\begin{align}
    S_\C[\Phi]=\int_\C\d^4 x\, \L[\Phi]\,,
\end{align}
where the subscript ``$\C$'' denotes that the integral over time is performed on the Keldysh contour while the integral over space is as usual.

The closed contour is composed of a forward branch (upper one in Fig.~\ref{fig:keldyshcontour}) and backward branch (lower one in Fig.~\ref{fig:keldyshcontour}), and defines the time ordering, $T_\C$. To distinguish the forward and backward branches, we can write the time variable on the forward branch as $t^+$ and on the backward branch as $t^-$. In an alternative convention, we use the common time variable but distinguish the fields on the forward and backward branches, $\Phi^+(t,\vec{x})=\Phi(t^+,\vec{x})$, $\Phi^-(t,\vec{x})=\Phi(t^-,\vec{x})$. For example, $S_{\C}[\Phi]$ can be written as $S[\Phi^+,\Phi^-]\equiv\int\d^4 x\, \left(\L[\Phi^+]-\L[\Phi^-]\right)=S[\Phi^+]-S[\Phi^-]$ where the minus sign is due to the opposite time direction in the lower branch of the original contour $\C$. Using the double-field notation, the integral is {\it not} performed on the closed contour anymore. 

We have four different types of propagators\footnote{Note that in the literature, some authors use a slightly different notation for the propagators: $\i\Delta_\phi=\langle \Phi\Phi\rangle_c$. \label{footnote3}}, namely
\begin{subequations}
\begin{align}
    &\Delta_\phi^{++}(x_1,x_2)=\langle T_\C \Phi^+(x_1)\Phi^+(x_2)\rangle_c=\langle T \Phi(x_1)\Phi(x_2)\rangle_c\,,\\[1mm]
    &\Delta^{-+}_\phi(x_1,x_2)\equiv \Delta^{>}_\phi(x_1,x_2)=\langle T_\C \Phi^-(x_1)\Phi^+(x_2)\rangle_c=\langle \Phi(x_1)\Phi(x_2)\rangle_c\,,\\[1mm]
    &\Delta_\phi^{+-}(x_1,x_2)\equiv \Delta^{<}_\phi(x_1,x_2)=\langle T_\C \Phi^+(x_1)\Phi^{-}(x_2)\rangle_c=\langle \Phi(x_2)\Phi(x_1)\rangle_c\,,\\[1mm]
    &\Delta_\phi^{--}(x_1,x_2)=\langle T_\C\Phi^-(x_1)\Phi^-(x_2)\rangle_c=\langle\overline{T}\Phi(x_1)\Phi(x_2)\rangle_c\,,
\end{align}
\end{subequations}
where $\overline{T}$ is the anti-time-ordering operator and $\Delta_\phi^{\gtrless}$ are {\it Wightman functions}. The retarded and advanced propagators are 
\begin{subequations}
\label{eq:retard-advanced-propagators}
\begin{align}
    &\Delta^r_\phi=\Delta^{++}_\phi -\Delta^<_\phi =\Delta^>_\phi -\Delta_\phi^{--}\,,\\[1mm]
    &\Delta^a_\phi =\Delta^{++}_\phi-\Delta^>_\phi =\Delta_\phi^<-\Delta_\phi^{--}\,.
\end{align} 
\end{subequations}
Their Hermitian and anti-Hermitian parts read
\begin{align}
    & \Delta^{\H}_\phi=-\frac{\i}{2} \left(\Delta_\phi^r+\Delta_\phi^a\right)\,,\\
    &\Delta_\phi^\A=\frac{1}{2} \left(\Delta_\phi^r-\Delta^a_\phi\right)=\frac{1}{2} \left(\Delta_\phi^>-\Delta_\phi^<\right)\,.
\end{align}
We also have two types of vertices: the $+$ and $-$ type. Based on these, one can construct Feynman rules for the perturbative expansion of generating functionals or effective actions. Due to the additional minus sign appearing in the integrals of the $-$ type fields, the Feynman rule for $-$ type vertices would differ from the corresponding $+$ type vertices by a minus sign. We also have two types of one-point functions $\varphi^+=\langle \Phi^+\rangle$ and $\varphi^-=\langle \Phi^-\rangle$. Since the physical condensate is unique, $\varphi^- =\varphi^+=\varphi$ should be imposed at the end of the calculation~\cite{Calzetta:1986cq}.\footnote{The condition $\varphi^+=\varphi^-$ can be understood as a consequence of the condition $J^+=J^-$ for the external sources, which must be imposed because the actual time axis is unique~\cite{Morikawa:1986rp}.} For instance, the 2PI effective action now takes the form $\Gamma_{\rm 2PI}[\varphi^a,\Delta^{ab}_\phi]$ where $a,b=\pm$. In this case, Eq.~\eqref{eq:kinetic-operator} becomes
\begin{align}
\label{eq:kinetic-operator-CTP}
    -\i\left.\frac{\delta^2 S[\Phi^+,\Phi^-]}{\delta\Phi^a(x_1)\delta\Phi^b(x_2)}\right|_{\varphi}=\delta^{(4)}(x_1-x_2) G^{ab,-1}_\phi\,.
\end{align}
The EoMs are given by
\begin{align}
        &\left.\frac{\delta \Gamma_{\rm 2PI}[\varphi,\Delta_\phi]}{\delta\varphi^+(x)}\right|_{\varphi^-=\varphi^+=\varphi}=0\,,\qquad \left.\frac{\delta \Gamma_{\rm 2PI}[\varphi,\Delta_\phi]}{\delta\Delta^{ab}_\phi(x_1,x_2)}\right|_{\varphi^-=\varphi^+=\varphi}=0\,.
\end{align}

\section{Non-local equations of motion}\label{sec:nonlocal-EoM}

We now apply the CTP formalism to our model. Substituting $\Phi=\varphi+\phi$ into Eq.~\eqref{eq:model}, the total Lagrangian can be written as
\begin{equation}
    \L=\bar{\L}[\varphi]+\hat{\L}[\phi,\chi;\varphi]\,.
\end{equation}
Here
\begin{subequations}
\begin{align}
\bar{\L}[\varphi] &= \frac{1}{2}(\partial_\mu\varphi)(\partial^\mu\varphi)-\frac{1}{2}m^2_\phi\varphi^2-\frac{\lambda_\phi}{4!}\varphi^4 \,, \\[1mm]
\hat{\L}[\phi,\chi;\varphi] &= \frac{1}{2} (\partial_\mu \phi) (\partial^\mu \phi)
+ \frac{1}{2} (\partial_\mu \chi) (\partial^\mu \chi)
-V(\phi,\chi;\varphi)\,, \label{eq:Sphichi}
\end{align}
\end{subequations}
where
\begin{align}
\label{eq:Vhphi}
V(\phi, \chi;\varphi) &=
\frac{1}{2} \left(m_\phi^2 + \frac{\lambda_{\phi}}{2}\varphi^2\right) \phi^2+\frac{1}{2} \left(m^2_\chi +  \frac{g}{2} \varphi^2\right) \chi^2+\frac{\lambda_\phi}{3!}\varphi\phi^3+\frac{g}{2}\varphi\phi \chi^2
+ \frac{\lambda_\phi}{4!} \phi^4\notag\\
&\quad+ \frac{\lambda_\chi}{4!} \chi^4 +\frac{g}{4}\phi^2 \chi^2 +({\rm linear\ terms\ in\ fluctuations})\,.
\end{align}
We are not concerned with the linear terms in fluctuations, e.g., $\lambda_\phi\varphi^3\phi$, $m_\phi \varphi \phi$ in our model, as they would give rise to one-particle-{\it reducible} (tadpole) diagrams, see e.g., Ref.~\cite{Peskin:1995ev}. For example, contracting the interaction term $\lambda_\phi \varphi^3\phi$ with any other term would give the following diagram
\begin{align}
\label{eq:tadpole}
\begin{tikzpicture}[baseline={0cm-0.5*height("$=$")}]
\draw[thick] (-0.5,0) -- (-0.6-0.3*0.707,-0.1-0.3*0.707) ;
\draw[thick] (-0.6-0.45*0.707,-0.1-0.45*0.707) circle (0.15) ;
\draw[thick] (-0.6-0.3*0.707,-0.1-0.3*0.707) -- (-0.6-0.6*0.707,-0.1-0.6*0.707) ;
\draw[thick] (-0.6-0.6*0.707,-0.1-0.3*0.707) -- (-0.6-0.3*0.707,-0.1-0.6*0.707) ;
\draw[thick] (-0.5,0) -- (-1,0) ;
\draw[thick] (-1.15,0) circle (0.15) ;
\draw[thick] (-1.15-0.1,0-0.1) -- (-1.15+0.1,0+0.1) ;
\draw[thick] (-1.15-0.1,0+0.1) -- (-1.15+0.1,0-0.1) ;
\draw[thick] (-0.5,0) -- (-0.6-0.3*0.707,0.1+0.3*0.707) ;
\draw[thick] (-0.6-0.45*0.707,0.1+0.45*0.707) circle (0.15) ;
\draw[thick] (-0.6-0.3*0.707,0.1+0.3*0.707) -- (-0.6-0.6*0.707,0.1+0.6*0.707) ;
\draw[thick] (-0.6-0.6*0.707,0.1+0.3*0.707) -- (-0.6-0.3*0.707,0.1+0.6*0.707) ;
\draw[thick] (-0.5,0) -- (0.25,0) ;
\filldraw (-0.5,0) circle (1.5pt) node {} ;
\draw [thick] (0.5,0) circle (0.25); 
\draw[thick] (0.5-0.17,0-0.17) -- (0.5+0.17,0+0.17) ;
\draw[thick] (0.5-0.17-0.07,0-0.17+0.05) -- (0.5+0.17-0.07,0+0.17+0.05) ;
\draw[thick] (0.5-0.17+0.07,0-0.17-0.05) -- (0.5+0.17+0.07,0+0.17-0.05);
\draw[thick] (0.5-0.17-0.07,0-0.17+0.05+0.1) -- (0.5+0.17-0.07-0.06,0+0.17+0.05+0.1-0.06) ;
\draw[thick] (0.5-0.17+0.1+0.06,0-0.17-0.07) -- (0.5+0.17+0.09,0+0.17-0.1-0.03) ;
\end{tikzpicture}\,,
\end{align}
where the blob represents an arbitrary tadpole diagram; see Appendix~\ref{app:Feynman} for the Feynman rules. The above diagram is separated into two disconnected parts when one cuts the $\phi$-propagator. However, we do allow terms linear in the condensate, $\varphi$, since the condensate acts in this case as an external classical source.

In principle, we need to construct a functional for the one-point function $\varphi$ and all the connected two-point functions, $\Gamma_{\rm 2PI}[\varphi,\Delta_\phi,\Delta_\Psi]$, where $\Delta_\Psi$ collectively denotes the connected two-point functions of the $\chi$ and SM fields, i.e., 
\begin{align}
    \Gamma_{\rm 2PI}[\varphi,\Delta_\phi,\Delta_\Psi] = &\, \bar{S}[\varphi^+]-\bar{S}[\varphi^-]+\frac{\i}{2}{\rm Tr}\ln \Delta_\phi^{-1}+\frac{\i}{2}{\rm Tr} \left[G_\phi^{-1}(\varphi)\Delta_\phi\right]+\frac{\i}{2}{\rm Tr}\ln \Delta_\chi^{-1}\notag\\
    &+\frac{\i}{2}{\rm Tr} \left[G_{\chi}^{-1}(\varphi)\Delta_\chi\right]+\sum_{\rm \Psi \, \neq \, \chi} a_{\Psi}\left[\i {\rm Tr}\ln \Delta_\Psi^{-1}+\i  {\rm Tr} \left(G_\Psi^{-1}\Delta_\Psi\right)\right]\notag\\
    &+\Gamma_2[\varphi,\Delta_\phi,\Delta_\Psi]\,,
\end{align}
where $\bar{S}[\varphi]=\int\d^4 x\, \bar{\L}[\varphi]$,  $a_{\Psi}=-1$ if $\Psi$ is a fermionic field, and $a_\Psi=1/2$ if it is a bosonic field. Now the trace is taken also for the Schwinger-Keldysh indices $a,b$. For instance,
\begin{align}
    {\rm Tr} \left[G^{-1}_\phi(\varphi)\Delta_\phi\right]=\sum_{a, \, b} \int\d^4 x\, \left.\left[G_\phi^{ab,-1}(\varphi(x)) \Delta_\phi^{ba}(x,y)\right]\right|_{y \, = \, x}\,.
\end{align}
If the two-point functions carry other indices, e.g., spinor or vector indices, one also needs to trace over these indices.

Above, the $\varphi$-dependence in the two fluctuation operators $G_\phi^{ab,-1}$ and $G_{\chi}^{ab,-1}$ have been indicated. Explicitly, they read 
\begin{subequations}
\begin{align}
   G_\phi^{ab,-1}(\varphi) &=\i c^{ab}\left[\Box+m_\phi^2+\frac{\lambda_\phi}{2}(\varphi^{a})^2\right]\,,\\
   \label{eq:G-inverse-A}
   G_{\chi}^{ab,-1}(\varphi) &=\i  c^{ab}\left[\Box+m_\chi^2+ \frac{g}{2}(\varphi^{a})^2\right]\,,
\end{align}
\end{subequations}
where $\Box=\partial_\mu\partial^\mu$, $c^{ab}$ is defined with $c^{++}=-c^{--}=1$ and $c^{+-}=c^{-+}=0$. The EoM for $\varphi$ with general $\Gamma_2[\varphi,\Delta_\phi,\Delta_\Psi]$ reads 
\begin{align}
    \left(\Box+m_\phi^2\right)\varphi(x)+\frac{\lambda_\phi}{6}\varphi^3(x)+\frac{\lambda_\phi}{2}\varphi(x)\Delta_\phi^{++}(x,x)+\frac{g}{2}\varphi(x) \Delta^{++}_\chi(x,x)-\left.\frac{\delta\Gamma_2}{\delta\varphi^+}\right|_{\varphi^+=\varphi^-=\varphi} =0\,. \label{eqn:first_condensate_eq}
\end{align}
The above equation can only be solved together with the complementary EoMs for other two-point functions. 
Let us discuss some of its properties. The first two terms, $\left(\Box+m_\phi^2\right)\varphi(x)+\lambda_\phi\varphi^3(x)/6$ are the classical, zero-temperature part for the EoM. The local term $g\varphi(x) \Delta^{++}_\chi(x,x)/2$ is linear in $\varphi$. Thus, when evaluated with a thermal distribution for $\chi$, this leads to a thermal mass term for the condensate. Similarly, the term $\lambda_\phi\varphi(x)\Delta_\phi^{++}(x,x)/2$ leads to an induced mass from the $\phi$ particles, and in general represents the local coupling between particles and condensate. The more complex term in this equation is $\delta\Gamma_2/\delta\varphi^+|_{\varphi^+=\varphi^-=\varphi}$ which is typically non-local. We evaluate this term below, giving an expression in terms of Feynman diagrams, and show how it leads to non-locality in the condensate EoM.

To introduce the EoM for the two-point function $\Delta_\phi^{ab}$, we first introduce the {\it $\phi$ particle self-energy}, 
\begin{align}
    \Pi_\phi^{ab}(x_1,x_2)&=-2(ab)\frac{\delta \Gamma_2[\varphi,\Delta_\phi,\Delta_\Psi]}{\delta \Delta_\phi^{ba}(x_2,x_1)}\,.
\end{align}
The EoM for the two-point functions generally reads as
\begin{align}
\label{eq:EoM-Deltaphi-general}
    &\left(-\Box-\mathfrak{m}_\phi^2\right)\Delta^{ab}_\phi(x_1,x_2)-\sum_c c\int\d^4 x_3\, \Pi_\phi^{ac}(x_1,x_3) \Delta^{cb}_\phi(x_3,x_2)=\i c^{ab} \delta^{(4)}(x_1-x_2)\,,
\end{align}
where $\mathfrak{m}^2_\phi=m_\phi^2+\lambda_\phi \varphi^2/2$, i.e., the effective particle mass in the presence of the condensate. 

Now we need to truncate $\Gamma_{\rm 2PI}$ to obtain an explicit expression of $\Gamma_2[\varphi,\Delta_\phi,\Delta_\Psi]$. We truncate $\Gamma_{\rm 2PI}$ at three loops so that the 2PI diagrams that have at least one DM propagator or condensate quantum are (symmetry factors are suppressed here)
\begin{align}
\label{eq:Gamma2}
\Gamma_2=-\i\Bigg(\,
&\begin{tikzpicture}[baseline={-0.025cm*height("$=$")}]
\draw[draw=black,thick] (-0.5,0) circle (0.5);
\filldraw (0,0) circle (1.5pt) node {} ;
\draw[draw=black,thick] (0.5,0) circle (0.5);
\end{tikzpicture}
\, +\,
\begin{tikzpicture}[baseline={-0.025cm*height("$=$")}]
\draw[draw=black,thick] (-0.5,0) circle (0.5);
\filldraw (0,0) circle (1.5pt) node {} ;
\draw[double,dashed] (0.5,0) circle (0.48);
\end{tikzpicture}
\, +\, 
\begin{tikzpicture}[baseline={-0.025cm*height("$=$")}]
\draw[draw=black,thick] (-0.5,0) circle (0.15) ;
\draw[draw=black,thick] (-0.5-0.12,0+0.12) -- (-0.5+0.12,0-0.12);
\draw[draw=black,thick] (-0.5-0.12,0-0.12) -- (-0.5+0.12,0+0.12);
\draw[draw=black,thick] (-0.35,0) -- (0.25,0);
\draw[draw=black,thick] (0.75,0) circle (0.5);
\filldraw (0.25,0) circle (1.5pt) node {} ;
\filldraw (1.25,0) circle (1.5pt) node {} ;
\draw[thick] (0.25,0) -- (1.25,0);
\draw[draw=black,thick] (1.25,0) -- (1.85,0);
\draw[draw=black,thick] (2,0) circle (0.15) ;
\draw[draw=black,thick] (2-0.12,0+0.12) -- (2+0.12,0-0.12);
\draw[draw=black,thick] (2-0.12,0-0.12) -- (2+0.12,0+0.12);
\end{tikzpicture}
\, +\,
\begin{tikzpicture}[baseline={-0.025cm*height("$=$")}]
\draw[draw=black,thick] (-0.5,0) circle (0.15) ;
\draw[draw=black,thick] (-0.5-0.12,0+0.12) -- (-0.5+0.12,0-0.12);
\draw[draw=black,thick] (-0.5-0.12,0-0.12) -- (-0.5+0.12,0+0.12);
\draw[draw=black,thick] (-0.35,0) -- (0.25,0);
\draw[double,dashed] (0.75,0) circle (0.48);
\filldraw (0.27,0) circle (1.5pt) node {} ;
\filldraw (1.23,0) circle (1.5pt) node {} ;
\draw[thick] (0.25,0) -- (1.25,0);
\draw[draw=black,thick] (1.25,0) -- (1.85,0);
\draw[draw=black,thick] (2,0) circle (0.15) ;
\draw[draw=black,thick] (2-0.12,0+0.12) -- (2+0.12,0-0.12);
\draw[draw=black,thick] (2-0.12,0-0.12) -- (2+0.12,0+0.12);
\end{tikzpicture}\notag 
\\
&\, +\,
\begin{tikzpicture}[baseline={-0.025cm*height("$=$")}]
\draw[thick] (0.75,0) circle (0.5);
\filldraw (0.25,0) circle (1.5pt) node {} ;
\filldraw (1.25,0) circle (1.5pt) node {} ;
\draw[thick] (1.25,-0.0) arc (30:150:0.57);
\draw[thick] (0.25,-0.0) arc (210:330:0.57);
\end{tikzpicture}
\, +\,
\begin{tikzpicture}[baseline={-0.025cm*height("$=$")}]
\draw[thick] (0.75,0) circle (0.5);
\filldraw (0.25,0) circle (1.5pt) node {} ;
\filldraw (1.25,0) circle (1.5pt) node {} ;
\draw[double,dashed] (1.25,-0.0) arc (30:150:0.57);
\draw[double,dashed] (0.25,-0.0) arc (210:330:0.57);
\end{tikzpicture}
\Bigg)
\,.
\end{align}

Taking the functional derivative of $\Gamma_2$ with respect to $\Delta_\phi$ amounts to cutting one $\phi$ propagator. Thus, 
$\Pi_\phi$ can be represented by the following diagrams 
\begin{align}
\label{eq:Pi-phi}
\Pi_\phi\sim \i \Bigg(\,
&\begin{tikzpicture}[baseline={-0.025cm*height("$=$")}]
\draw[black, thick] (0,0.25) circle (0.5);  
\filldraw (0,-0.25) circle (1.5pt) node {} ;
\end{tikzpicture}
\, +\, 
\begin{tikzpicture}[baseline={-0.025cm*height("$=$")}]
\draw[double, dashed] (0,0.25) circle (0.5);  
\filldraw (0,-0.25) circle (1.5pt) node {} ;
\end{tikzpicture}
\,+\,
\begin{tikzpicture}[baseline={-0.025cm*height("$=$")}]
\draw[draw=black,thick] (0.25,-0.6) circle (0.15) ;
\draw[draw=black,thick] (-0.5-0.12+0.75,0+0.12-0.6) -- (-0.5+0.12+0.75,0-0.12-0.6);
\draw[draw=black,thick] (-0.5-0.12+0.75,0-0.12-0.6) -- (-0.5+0.12+0.75,0+0.12-0.6);
\draw[draw=black,thick] (0.25,0) -- (0.25,-0.45);
\draw[draw=black,thick] (0.75,0) circle (0.5);
\filldraw (0.25,0) circle (1.5pt) node {} ;
\filldraw (1.25,0) circle (1.5pt) node {} ;
\draw[draw=black,thick] (1.25,-0.6) circle (0.15) ;
\draw[draw=black,thick] (-0.5-0.12+0.75+1,0+0.12-0.6) -- (-0.5+0.12+0.75+1,0-0.12-0.6);
\draw[draw=black,thick] (-0.5-0.12+0.75+1,0-0.12-0.6) -- (-0.5+0.12+0.75+1,0+0.12-0.6);
\draw[draw=black,thick] (1.25,0) -- (1.25,-0.45);
\end{tikzpicture}
\,+\,
\begin{tikzpicture}[baseline={-0.025cm*height("$=$")}]
\draw[draw=black,thick] (0.25,-0.6) circle (0.15) ;
\draw[draw=black,thick] (-0.5-0.12+0.75,0+0.12-0.6) -- (-0.5+0.12+0.75,0-0.12-0.6);
\draw[draw=black,thick] (-0.5-0.12+0.75,0-0.12-0.6) -- (-0.5+0.12+0.75,0+0.12-0.6);
\draw[draw=black,thick] (0.25,0) -- (0.25,-0.45);
\draw[double,dashed] (0.75,0) circle (0.5);
\filldraw (0.25,0) circle (1.5pt) node {} ;
\filldraw (1.25,0) circle (1.5pt) node {} ;
\draw[draw=black,thick] (1.25,-0.6) circle (0.15) ;
\draw[draw=black,thick] (-0.5-0.12+0.75+1,0+0.12-0.6) -- (-0.5+0.12+0.75+1,0-0.12-0.6);
\draw[draw=black,thick] (-0.5-0.12+0.75+1,0-0.12-0.6) -- (-0.5+0.12+0.75+1,0+0.12-0.6);
\draw[draw=black,thick] (1.25,0) -- (1.25,-0.45);
\end{tikzpicture}
\, +\,
\begin{tikzpicture}[baseline={-0.025cm*height("$=$")}]
\draw[draw=black,thick] (0.75,0) circle (0.5);
\filldraw (0.25,0) circle (1.5pt) node {} ;
\filldraw (1.25,0) circle (1.5pt) node {} ;
\draw[thick] (0.25,0) -- (1.25,0);
\end{tikzpicture}
\,+\,
\begin{tikzpicture}[baseline={-0.025cm*height("$=$")}]
\draw[double,dashed] (0.75,0) circle (0.48);
\filldraw (0.25,0) circle (1.5pt) node {} ;
\filldraw (1.25,0) circle (1.5pt) node {} ;
\draw[thick] (0.25,0) -- (1.25,0);
\end{tikzpicture}
\Bigg) 
\,.
\end{align}
Explicitly, it reads 
\begin{align}
\label{eq:Pi-phi-ex}
    \Pi_\phi^{ab}(x_1,x_2)=&\frac{\lambda_\phi}{2} a \delta^{ab}\Delta_\phi^{ab}(x_1,x_2)\delta^{(4)}(x_1-x_2)+ \frac{g}{2} a \delta^{ab} \Delta_\chi^{ab}(x_1,x_2)\delta^{(4)}(x_1-x_2) \notag\\
    &-\frac{\i\lambda_\phi^2}{2} \varphi^a(x_1)[\Delta_\phi^{ab}(x_1,x_2)]^2\varphi^b(x_2) -\frac{\i g^2}{2} \varphi^a(x_1)[\Delta_\chi^{ab}(x_1,x_2)]^2\varphi^b(x_2)\notag\\
    &-\frac{\i\lambda_\phi^2}{3!} [\Delta_\phi^{ab}(x_1,x_2)]^3-\frac{\i g^2}{2} \Delta_\phi^{ab}(x_1,x_2)[\Delta_\chi^{ab}(x_1,x_2)]^2 \,.  
\end{align}
As for the two-point functions $\Delta_{\phi,\chi}$, we also introduce the notation
\begin{align}
    \Pi^{>}_\phi\equiv \Pi^{-+}_\phi\,, \quad \Pi^{<}_\phi\equiv \Pi^{+-}_\phi\,,
\end{align}
that will be used later. Note that the local terms in Eq.~\eqref{eq:Pi-phi-ex} (the first two terms) vanish for $\Pi_\phi^{<,>}$ due to the presence of $\delta^{ab}$ in those terms.

Taking the functional derivative of $\Gamma_2$ with respect to $\varphi$ amounts to cutting one condensate quantum. Thus, diagrammatically, one has
\begin{align}
\label{eq:Sigma-diagram}
\frac{\delta\Gamma_2}{\delta\varphi}\sim -\i \Bigg(\,
\begin{tikzpicture}[baseline={-0.025cm*height("$=$")}]
\draw[black,thick] (0.75,0) circle (0.5);
\filldraw (0.27,0) circle (1.5pt) node {} ;
\filldraw (1.23,0) circle (1.5pt) node {} ;
\draw[thick] (0.25,0) -- (1.25,0);
\draw[draw=black,thick] (1.25,0) -- (1.85,0);
\draw[draw=black,thick] (2,0) circle (0.15) ;
\draw[draw=black,thick] (2-0.12,0+0.12) -- (2+0.12,0-0.12);
\draw[draw=black,thick] (2-0.12,0-0.12) -- (2+0.12,0+0.12);
\end{tikzpicture}
\,+\,
\begin{tikzpicture}[baseline={-0.025cm*height("$=$")}]
\draw[double,dashed] (0.75,0) circle (0.48);
\filldraw (0.27,0) circle (1.5pt) node {} ;
\filldraw (1.23,0) circle (1.5pt) node {} ;
\draw[thick] (0.25,0) -- (1.25,0);
\draw[draw=black,thick] (1.25,0) -- (1.85,0);
\draw[draw=black,thick] (2,0) circle (0.15) ;
\draw[draw=black,thick] (2-0.12,0+0.12) -- (2+0.12,0-0.12);
\draw[draw=black,thick] (2-0.12,0-0.12) -- (2+0.12,0+0.12);
\end{tikzpicture}
\Bigg) 
\,.
\end{align}
Usually, $-\delta^2\Gamma_2/\delta\varphi(x)\delta\varphi(y)$ is also called the self-energy. This self-energy is responsible for the condensate dynamics and therefore we refer to it as the {\it $\varphi$ condensate self-energy}. Diagrammatically, we have
\begin{align}
\label{eq:Pi-varphi}
    -(ab)\frac{\delta^2\Gamma_2}{\delta\varphi^a(x)\delta\varphi^b(y)}\equiv\Pi^{ab}_\varphi \sim 
    \i\Bigg(\,
\begin{tikzpicture}[baseline={-0.025cm*height("$=$")}]
\draw[black,thick] (0.75,0) circle (0.5);
\filldraw (0.27,0) circle (1.5pt) node {} ;
\filldraw (1.23,0) circle (1.5pt) node {} ;
\draw[thick] (0.25,0) -- (1.25,0);
\end{tikzpicture}
\,+\,
\begin{tikzpicture}[baseline={-0.025cm*height("$=$")}]
\draw[double,dashed] (0.75,0) circle (0.48);
\filldraw (0.27,0) circle (1.5pt) node {} ;
\filldraw (1.23,0) circle (1.5pt) node {} ;
\draw[thick] (0.25,0) -- (1.25,0);
\end{tikzpicture}
\Bigg) \,.
\end{align}
Note the different subscripts in $\Pi_\phi$ and $\Pi_\varphi$. Explicitly, $\Pi^{ab}_\varphi$ reads
\begin{align}
\label{eqn:condensate_self_energy}
    \Pi^{ab}_\varphi(x,y)=-\frac{\i \lambda_\phi^2}{3!} \left(\Delta_\phi^{ab}(x,y)\right)^3 -\frac{\i g^2}{2} \Delta_\phi^{ab}(x,y) \left(\Delta_\chi^{ab}(x,y)\right)^2 \,.
\end{align}

Apparently, the $\varphi$ condensate self-energy $\Pi_\varphi$ is not necessarily the same as the $\phi$ particle self-energy $\Pi_\phi$. Although the two diagrams in Eq.~\eqref{eq:Pi-varphi} are the same as the last two diagrams in Eq.~\eqref{eq:Pi-phi}, the processes they describe actually correspond to the third and fourth diagrams in Eq.~\eqref{eq:Pi-phi}. This can be easily understood from the fact that these diagrams, respectively, come from the same diagrams in $\Gamma_2$, the third and fourth diagrams in Eq.~\eqref{eq:Gamma2}.

Thus, we find the explicit relationship for the functional derivative of the effective action:
\begin{align}
    -\left.\frac{\delta\Gamma_2[\varphi,\Delta_\phi,\Delta_\Phi]}{\delta\varphi^+}\right|_{\varphi^+=\varphi^-=\varphi}=\int\d^4 y\, \Pi^{\rm r}_\varphi(x,y)\varphi(y)\,, \label{eqn:dGam2_dphi+}
\end{align}
where the retarded $\varphi$-condensate self-energy reads
\begin{align}\label{eqn:retarded_condensate_self_energy}
    \Pi^{\rm r}_\varphi(x,y)=\Pi_\varphi^{++}-\Pi_\varphi^{+-}=&-\frac{\i \lambda_\phi^2}{3!} \left[(\Delta_\phi^{++}(x,y))^3-(\Delta_\phi^{+-}(x,y))^3\right]  \notag\\
    &-\frac{\i g^2}{2} \left[\Delta_\phi^{++}(x,y) (\Delta_\chi^{++}(x,y))^2-\Delta_\phi^{++}(x,y)(\Delta_\phi^{+-}(x,y))^2\right] \,.
\end{align}
Using Eqs.~\eqref{eqn:dGam2_dphi+} and \eqref{eqn:retarded_condensate_self_energy}, we can evaluate the last term in the condensate EoM, Eq.~\eqref{eqn:first_condensate_eq}. This term, as noted earlier, involves an integral over the second argument of $\Pi_\varphi^r$. The integral over the retarded self-energy covers the entire past lightcone, introducing a non-locality and memory into the EoM.

Now we further develop the EoM for the two-point function, Eq.~\eqref{eq:EoM-Deltaphi-general}. The condensate enters only through $\Pi_\phi^{ab}$. For the moment, we keep $\Pi_\phi^{ab}$ general and introduce the standard localization procedure (see e.g., Ref.~\cite{Prokopec:2003pj}). In section~\ref{sec:Boltzmann}, we provide our specific treatment to $\Pi^{ab}_\phi$ in the presence of oscillating condensates.

For later convenience, the retarded and advanced self-energies are defined by Eq.~\eqref{eq:retard-advanced-propagators} with $\Delta_\phi$ replaced with $\Pi_\phi$. The Hermitian and non-Hermitian parts read, however, differently:\footnote{Had we used the convention $\i \Delta_\phi=\langle \Phi\Phi\rangle_c$ mentioned in footnote~\ref{footnote3}, the expressions of $\Delta_\phi^{\A,\H}$ would take the same form as Eqs.~\eqref{eq:Pi-H-A}.}
\begin{subequations}
\label{eq:Pi-H-A}
\begin{align}
&\Pi^\H_\phi=\frac{1}{2} \left(\Pi_\phi^r+\Pi_\phi^a\right)\,,\\
&\Pi^\A_\phi=\frac{\i}{2}\left(\Pi_\phi^r-\Pi_\phi^a\right)\,.
\end{align}
\end{subequations}
We now consider the EoMs for $\Delta^>_\phi$ and $\Delta^<_\phi$. They can be compactly written as (c.f. Eq.~\eqref{eq:EoM-Deltaphi-general})
\begin{align}
\label{eq:EoM-Delta><2}
    \left(-\Box - \mathfrak{m}_\phi^2\right) \Delta_\phi^{<,>}-\Pi^r_\phi \odot \Delta_\phi^{<,>} = \Pi^{<,>}_\phi \odot \Delta_\phi^a\,,
\end{align}
where $\odot$ stands for integration over the intermediate variable. Note that the right-hand side of Eq.~\eqref{eq:EoM-Deltaphi-general} vanishes due to $c^{+-}=c^{-+}=0$. In terms of the Hermitian and non-Hermitian parts of the retarded and advanced propagators/self-energies, the above equation reads 
\begin{align}
\label{eq:EoM-Delta><1}
    \left(-\Box -\mathfrak{m}_\phi^2\right) \Delta_\phi^{<,>}-\Pi^\H_\phi \odot \Delta_\phi^{<,>} -\i\, \Pi^{<,>}_\phi \odot \Delta_\phi^\H =C_\phi\,,
\end{align}
where 
\begin{align}
    C_\phi= \left(\Pi^>_\phi \odot \Delta_\phi^< -\Pi^<_\phi \odot \Delta_\phi^> \right)\,,
\end{align}
which is suggestively labelled $C_\phi$. As we will see below, this gives rise to the collision term in the Boltzmann equation.

Eq.~\eqref{eq:EoM-Delta><1}, via the $\phi$ particle self-energy $\Pi_\phi$, involves interactions between the $\phi$ particles, $\chi$ particles, and $\varphi$ condensate. The integration in the $\odot$ operator makes this non-local, and non-Markovian. In the next section, we introduce a method to obtain a familiar Markovian EoMs.

\section{Markovianization}\label{sec:markovianize}

\subsection{The two-point function:~Wigner transform and gradient expansion}

To localize/Markovianize Eq.~\eqref{eq:EoM-Delta><1}, one introduces the Wigner transform. For instance, the Wigner transform of $\Delta_\phi(x_1,x_2)$ is:
\begin{align}
    \overbar{\Delta}_\phi(k,x)=\int\d^4 r \, \e^{\i k \cdot r} \Delta_\phi\left(x+\frac{r}{2},\,x-\frac{r}{2}\right)\,,
\end{align}
where $x=(x_1+x_2)/2$ and $r=x_1-x_2$. We put a bar on the Wigner-transformed two-point functions. Here $x$ and $r$ can be viewed as the macroscopic and microscopic coordinates, respectively. Thus, the Wigner transform is simply the Fourier transform for the microscopic coordinate. The Wigner transform separates macroscopic evolution in real space of the ``center of mass'' coordinate $x$, from microscopic processes characterized by the conjugate momentum $k$ of the relative coordinate $r$.

In Wigner space, we have 
\begin{align}
&\left[\overbar{\Delta}^{<,>}_\phi(k,x)\right]^\dagger =\overbar{\Delta}^{<,>}_\phi(k,x)\,.
\end{align}
For the convolutions, there is the following identity~\cite{Groenewold:1946kp,Moyal:1949sk}
\begin{align}
    \int\d^4 (x_1-x_2)\, \e^{\i k\cdot (x_1-x_2)}\int\d^4 x_3\, A(x_1,x_3) B(x_3,x_2)=\e^{-\i \diamond}\{ \overbar{A}(k,x) \}\{\overbar{B}(k,x)\}\,,
\end{align}
where the diamond operator is defined as
\begin{align}
    \diamond \{ \overbar{A}(k,x) \}\{\overbar{B}(k,x)\}=\frac{1}{2}\left(\frac{\partial \overbar{A}(k,x)}{\partial x^\mu} \frac{\partial \overbar{B}(k,x)}{\partial k_\mu}-\frac{\partial \overbar{A}(k,x)}{\partial k_\mu} \frac{\partial \overbar{B}(k,x)}{\partial x^\mu}\right)\,.
\end{align}
Thus, in Wigner space, Eq.~\eqref{eq:EoM-Delta><1} can be written as
\begin{align}
\label{eq:EoM-Delta><3}
    \left(k^2-\frac{1}{4} \partial_x^2 +\i k\cdot \partial_x -\mathfrak{m}_\phi^2 \e^{-\frac{\i}{2}  \overleftarrow{\partial_{x}}\cdot\partial_{k} } \right) \overbar{\Delta}^{<,>}_\phi - \e^{-\i\diamond} \{\overbar{\Pi}^\H_\phi\}\{\overbar{\Delta}_\phi^{<,>}\}-\i \e^{-\i\diamond} \{\overbar{\Pi}^{<,>}_\phi\}\{\overbar{\Delta}^\H_\phi\}=\overbar{C}_\phi\,,
\end{align}
where
\begin{align}
    \overbar{C}_\phi=\frac{1}{2}\e^{-\i\diamond} \left(\{\overbar{\Pi}_\phi^>\}\{\overbar{\Delta}_\phi^<\}-\{\overbar{\Pi}_\phi^<\}\{\overbar{\Delta}_\phi^>\}\right)\,.
\end{align}
The non-locality of Eq.~\eqref{eq:EoM-Delta><2} has been transformed into infinitely many derivatives in Eq.~\eqref{eq:EoM-Delta><3}, which appear by Taylor expanding the exponential of the diamond operator, leading to a power series expansion in gradients that can be controlled order-by-order.

We shall truncate Eq.~\eqref{eq:EoM-Delta><3} at the first order in the gradient expansion. The Hermitian part gives the constraint equation and the anti-Hermitian part gives the kinetic equation. The latter reads
\begin{align}
\label{eq:kinetic-eq}
    \left[k_\mu \partial_{(x)}^\mu +\frac{1}{2}\left(\partial^{(x)}_\mu M_\phi^2\right)\partial^\mu_{(k)} \right] \overbar{\Delta}_\phi^{<,>}=-\frac{\i}{2} \left(\overbar{\Pi}_\phi^> \overbar{\Delta}_\phi^< -\overbar{\Pi}_\phi^< \overbar{\Delta}_\phi^>\right)\,,
\end{align}
where 
\begin{align}
    M_\phi^2=\mathfrak{m}_\phi^2+\overbar{\Pi}^\H_\phi\,.
\end{align}
Thus, the Hermitian part of the $\phi$ self-energy contributes to the effective mass at leading order in the gradient expansion. Similarly, we denote $M^2_\chi\equiv \mathfrak{m}_\chi^2+\overbar{\Pi}^\H_\chi\equiv m_\chi^2+g\varphi^2/2+\overbar{\Pi}_\chi^\H$. The gradient expansion is based on the assumption of small variations of $M_\phi$, $\overbar{\Delta}_{\phi,\,\chi}$ with respect to the macroscopic coordinate $x$ compared to the typical microscopic scale of the corresponding species in the plasma. In our case, namely a spatially homogeneous but oscillating background, it amounts to the following criterion 
\begin{align}
\label{eq:condition-gradEx}
  \frac{\dot{M}_{\phi,\chi}}{M^2_{\phi,\chi}}\ll 1\,,
\end{align}
where a dot denotes the time derivative.
Eq.~\eqref{eq:condition-gradEx} should be a condition when applying the derived Boltzmann equations to phenomenological studies. In particular, this limit applies for the QCD axion temperature-dependent mass once $m_a(T)\gg H$, since $\dot{m}_a/m_a\sim \dot{T}/T\sim H$.

To see that Eq.~\eqref{eq:kinetic-eq} can lead to the relativistic Boltzmann equation, let us consider, for instance, the sixth diagram (hence the subscript ``(6)'' in the following equation) in Eq.~\eqref{eq:Pi-phi} such that
\begin{align}
\label{eq:Piphi-sunset-phi-chi}
    \Pi_{\phi,(6)}^{ab}(x_1,x_2)=-\frac{\i g^2}{2}\Delta_\phi^{ab}(x_1,x_2) [\Delta_\chi^{ab}(x_1,x_2)]^2\,.
\end{align}
Taking the Wigner transform of the above equation gives
\begin{align}
\label{eq:Pi-phi-6}
    \overbar{\Pi}^{<,>}_{\phi,(6)}(k,x)=-\frac{\i g^2}{2} &\int\prod_{i=1,\,2,\,3} \frac{\d^4 k_i}{(2\pi)^4} (2\pi)^4
    \delta(k-k_1-k_2-k_3) \notag \\
    &\qquad \quad \quad \times \overbar{\Delta}_\phi^{<,>}(k_1,x)\overbar{\Delta}_\chi^{<,>}(k_2,x)\overbar{\Delta}_\chi^{<,>}(k_3,x).
\end{align}

The key step to arrive at the Boltzmann equation is to consider the on-shell limit, in which the Wightman functions read~\cite{Prokopec:2003pj} 
\begin{subequations}
\label{eq:on-shell-limit}
    \begin{align}
    \overbar{\Delta}_{\phi/\chi}^{<}(k,x)&=2\pi \delta(k^2-M^2_{\phi,\chi})\,{\sign} (k_0)f_{\phi/\chi}(k,x)\,,\\[1mm]
    \overbar{\Delta}_{\phi/\chi}^>(k,x)&=2\pi \delta (k^2-M^2_{\phi,\chi})\,{\sign}(k_0)(1+f_{\phi/\chi}(k,x))\,.
\end{align}
\end{subequations}
The Wightman functions are rightly distributions, i.e., defined  under integrals. Thus, they have the right character to act as the distribution function, $f$.

Here we give some comments on the on-shell limit. To the leading order in gradients, the retarded and advanced propagators take the following form~\cite{Prokopec:2003pj}
\begin{align}
    \overbar{\Delta}_{\phi/\chi}^{r,a}=\frac{1}{\Omega_{\phi/\chi}^2\pm \i \Gamma_{\phi/\chi}}\,,
\end{align}
where $\Omega_{\phi,\chi}^2=k^2-\mathfrak{m}_{\phi,\chi}^2-\overbar{\Pi}_{\phi,\chi}^\H$ and $\Gamma_{\phi,\chi}\equiv \overbar{\Pi}^\A_{\phi,\chi}$. Physically, $\overbar{\Pi}^\H$ can be viewed as ($k$-dependent) corrections to the mass-squared, whereas $\overbar{\Pi}^\A$ represents finite widths of the collective excitations in the plasma. The on-shell limit is the limit $\Gamma\rightarrow 0$ under which the quasi-particle picture can be used. However, taking this limit should be understood properly. As $\overbar{\Pi}^\H$ and $\overbar{\Pi}^\A$ are controlled by the same coupling constants, they actually acquire contributions at the same order in the coupling constant. This means that in the strict limit $\Gamma\rightarrow 0$, one also has $\overbar{\Pi}^\H\rightarrow 0$. Moreover, the strict limit $\Gamma\rightarrow 0$ is implemented only for free theories, meaning that taking the limit $\Gamma\rightarrow 0$ would force us to neglect the collision terms. Therefore, substituting Eqs.~\eqref{eq:on-shell-limit} into the kinetic equation while keeping the collision terms on the right-hand side should be understood as an expansion of the kinetic equation in $\Gamma$. The self-consistency is assured by the fact that the dependence on $\Gamma$ of the left-hand side starts from $\O(\Gamma^2)$ while the right-hand side is of order $\O(\Gamma)$. For more discussions on this, see Ref.~\cite{Prokopec:2003pj}.

We now use the relations 
\begin{align}
\label{eq:relation1}
    \delta(k^2-M^2_{\phi/\chi})\,{\sign}(k_0)=\frac{\delta(k_0-E_\veck^{(\phi/\chi)})}{2 E_\veck^{(\phi/\chi)}}-\frac{\delta(k_0+E_\veck^{(\phi/\chi)})}{2E_\veck^{(\phi/\chi)}}
\end{align}
and\footnote{This relation can be derived from the $\C$ (charge conjugate) symmetry of the theory~\cite{Prokopec:2003pj}.}
\begin{align}
\label{eq:relation2}
    f_{\phi/\chi}(-k,x)=- \left(1+f_{\phi/\chi}(k,x) \right)\,.
\end{align}
One can recognize that $-\i/2 (\overbar{\Pi}_{\phi,(6)}^> \overbar{\Delta}_\phi^< -\overbar{\Pi}_{\phi,(6)}^< \overbar{\Delta}_\phi^>)$ gives precisely the collision terms for all the possible processes (provided the kinetic conditions can be satisfied) with two $\phi$ particles and two $\chi$ particles. One can understand these processes by the cutting rules at finite temperature~\cite{Cutkosky:1960sp,Weldon:1983jn,Kobes:1985kc,Kobes:1986za,Landshoff:1996ta,Gelis:1997zv,Bedaque:1996af} by which one cuts, in the example we are discussing, the last diagram in Eq.~\eqref{eq:Pi-phi}, and allows all combinations of incoming and outgoing particles. In Appendix~\ref{app:cutting}, we illustrate the cutting rules in more detail.

Thus, Eq.~\eqref{eq:kinetic-eq} reads in the on-shell limit in general as the relativistic Boltzmann equation:
\begin{align}
\label{eq:rel-Boltzmann-particle}
    \frac{1}{k^0}\left[k_\mu \partial_{(x)}^\mu +\frac{1}{2}\left(\partial^{(x)}_\mu M_\phi^2\right)\partial^\mu_{(k)} \right] f_\phi(k,x)= \C[f_\phi,f_i]\,,
\end{align}
where\footnote{We have written the collision term in the general form that is also valid for fermions.}
\begin{align}
\label{eq:collision-term}
    \C[f_\phi,f_i]=&\sum_{\rm all\ possible\ processes}\Bigg[\frac{1}{2 k^0}\int \prod_i \frac{\d^3 \veck_i}{(2\pi)^3 \, 2 k^0_i} (2\pi)^4 \delta^{(4)}(k+k_{A1}+\cdots-k_{B1}-\cdots)\notag\\
    &\hspace{5mm} \times \{(1\pm f_\phi)(1\pm f_{A1})\cdots f_{B1}\cdots-f_\phi f_{A1}\cdots (1\pm f_{B1})\cdots\} \, |\M_{\phi A_1\cdots\rightarrow B_1\cdots}|^2\Bigg]\,.
\end{align}
The index $i$ runs over particles $\{A_1,...,B_1,...\}$ corresponding to all the propagators appearing in the diagram of $\Pi_\phi$ under study. The upper and lower signs in $\pm$ correspond to bosons and fermions, respectively.

Note that in Eq.~\eqref{eq:rel-Boltzmann-particle}, all particles are on-shell, meaning that $k_i^0=\sqrt{M_i^2+\veck^2_i}$ ($M_i$ is the mass of the particle labeled by $i$), and $k^0=\sqrt{M_\phi^2+\veck^2}$. However, on the left-hand side, one should impose the on-shell condition $k^2=M_\phi^2$ only after the derivatives of $f_\phi$ with respect to the four-momentum are taken. This should be clear if one does the exercise of deriving Eq.~\eqref{eq:rel-Boltzmann-particle} from~\eqref{eq:kinetic-eq} as described above.

When a self-energy diagram (e.g., the third and fourth diagrams in Eq.~\eqref{eq:Pi-phi}) involves the condensate, the general form of Eq.~\eqref{eq:rel-Boltzmann-particle} should still hold. However, the presence of the condensate can modify the invariant transition amplitude $\M$. Below, we shall see that the background field can gain or lose condensate quanta from the interaction with ordinary particles or simply modify the vertex without any energy transfer between the condensate and ordinary particles. To discuss these processes, we now turn to the discussion of the condensate EoM. 

\subsection{The condensate: small-field expansion and multiple-scale analysis}
\label{sec:small-field}

\paragraph{Small-field expansion in $\Delta_\chi[\varphi]$} As mentioned earlier, we assume that the $\chi$ field and all the SM fields are in thermal equilibrium. Thus, we are not interested in the EoMs of the two-point functions for those fields; they are approximated by the free thermal equilibrium propagators. However, there is a problem with this simplification. The condensate explicitly modifies the dispersion relation of the $\chi$ field; one in principle needs to study and solve the EoM for $\Delta_\chi$. The solution would have a dependence on $\varphi$, i.e., $\Delta_\chi =\Delta_\chi[\varphi]$.
If the modification in the dispersion relation due to the $\varphi$-dependent mass term is significant, the dynamics of $\chi$ particles would be non-perturbative. Below, we shall assume that the  $\varphi$-dependent mass term for the $\chi$ field is small compared to its $\varphi$-independent mass so that one can perform the small-field expansion in the EoM for $\Delta_\chi$ or effectively in the solution $\Delta_\chi[\varphi]$. 

At the leading order, $\mathbf{\Delta}_\chi\equiv \Delta_\chi[\varphi=0]$ is given by the free thermal equilibrium propagator but with thermal-mass corrections taken into account~\cite{Ai:2021gtg}. Specifically, for the Wightman functions in Wigner space, we have
\begin{subequations}
\label{eq:Deltachi-varphi0}
    \begin{align}
    \overbar{\mathbf{\Delta}}^{<}_{\chi}(k) &=2\pi \delta(k^2-\widetilde{M}^2_{\chi})\,{\sign} (k_0)f^{\rm eq}_{\chi}(k_0)\,,\\
    \overbar{\mathbf{\Delta}}^{>}_{\chi}(k) &=2\pi \delta (k^2-\widetilde{M}^2_{\chi})\,{\sign}(k_0)(1+f^{\rm eq}_{\chi}(k_0))\,,
\end{align}
\end{subequations}
where $\widetilde{M}^2_\chi=m_\chi^2+\Pi_\chi^\H$ and $f^{\rm eq}_\chi$ is the equilibrium Boltzmann distribution function.

Our strategy for truncation in the small-field expansion is as follows. We will simply replace the $\varphi$-dependent $\chi$ propagators in $\Gamma_2$ by the leading-order results  $\mathbf{\Delta}_\chi$. However, for the term with $\Delta_\chi^{++}(x,x)$ in the condensate EoM, the small-field expansion at the next-to-leading order generates new processes that we would like to take into account. Diagrammatically, 
\begin{align}
\begin{tikzpicture}[baseline={-0.025cm*height("$=$")}]
\draw[double,dashed] (0.75,0) circle (0.48);
\filldraw (0.26,0) circle (1.5pt) node {} ;
\end{tikzpicture}
=
\begin{tikzpicture}[baseline={-0.025cm*height("$=$")}]
\draw[dashed] (0.75,0) circle (0.48);
\filldraw (0.26,0) circle (1.5pt) node {} ;
\end{tikzpicture}\,
+
\begin{tikzpicture}[baseline={-0.025cm*height("$=$")}]
\draw[dashed] (0.75,0) circle (0.48);
\filldraw (0.26,0) circle (1.5pt) node {} ;
\filldraw (1.24,0) circle (1.5pt) node {};
\draw[thick,black] (1.24,0) -- (1.6,0.36) ;
\draw[thick,black] (1.24,0) -- (1.6,-0.36);
\draw[thick,black] (1.64+0.075,0.4+0.075) circle (0.15);
\draw[thick,black] (1.6,0.36) -- (1.6+0.22,0.36+0.22);
\draw[thick,black] (1.605,0.6) -- (1.605+0.22,0.6-0.22);
\draw[thick,black] (1.64+0.075,-0.4-0.075) circle (0.15);
\draw[thick,black] (1.6,-0.36) -- (1.6+0.22,-0.36-0.22);
\draw[thick,black] (1.605,-0.6) -- (1.605+0.22,-0.6+0.22);
\end{tikzpicture}\,
+
\O(\varphi^4)\,,
\end{align}
where a single dashed line denotes the leading-order result, $\mathbf{\Delta}_\chi$. The second term on the right-hand side describes the dissipative process $\varphi\varphi\leftrightarrow \chi\chi$  (when this process is kinematically allowed). The condensate EoM can then be written as
\begin{align}
\label{eq:eom-varpphi2}
    \left(\Box+\overbar{M_\phi}^2\right)\varphi(x)+\frac{\lambda_\phi}{6}\varphi^3(x) + \int \d^4 x'\, \Pi_\varphi^r(x,x')\varphi(x') +\frac{\varphi(x)}{6}\int \d^4 x'\, V_\varphi^r(x-x') \varphi^2(x') =0\,,
\end{align}
where 
\begin{align}
\label{eq:M2}
    \overbar{M_\phi}^2(x) &= m_\phi^2+\frac{g}{2}\mathbf{\Delta}_\chi^{++} (0)+\frac{\lambda_\phi}{2}\Delta_\phi^{++}(x,x)\,,\\
    V_\varphi^r(x-x') &=-\frac{3\i g^2}{2} \left[\left(\mathbf{\Delta}_\chi^{++}(x-x')\right)^2-\left(\mathbf{\Delta}_\chi^{+-}(x-x')\right)\right]\,.
\end{align}
Here $V_\varphi^r$ is called the retarded proper four-vertex function, and $\overbar{M_\phi}$ is the leading-order frequency of the oscillation in multiple-scale analysis. Above, we have used the space-time-translation invariance for the thermal propagator, $\mathbf{\Delta}_\chi (x,x')=\mathbf{\Delta}_\chi(x-x')$. For $\Delta_\phi$, such property is absent because of the explicit $\varphi$-dependent mass for $\phi$ particles and because they are not assumed to be in thermal equilibrium (which is the case of interest during freeze-in DM production).

In this work, for simplicity, we are interested in homogeneous scalar backgrounds, such as one obtained for the axion when spontaneous symmetry breaking occurs during inflation. Taking into account also the term $3H\dot{\varphi}$ due to the expansion of the Universe, the EoM is: 
\begin{align}
\label{eq:eom-varpphi3}
    \ddot{\varphi}(t)+3H\dot{\varphi}(t)+\overbar{M_\phi}^2\varphi(t)+\frac{\lambda_\phi}{6}\varphi^3(t) + \int \d t'\, \pi_\varphi^r(t,t')\varphi(t') +\frac{\varphi(t)}{6}\int \d t'\, v_\varphi^r(t-t') \varphi^2(t') =0\,,
\end{align}
where 
\begin{align}
\pi_\varphi^r(t,t') &= \int \d^{3}(\vecx-\vecx') \, \Pi_\varphi^r(t,t';\vecx-\vecx') \,, \\
v_\varphi^r(t-t') &= \int \d^{3}(\vecx-\vecx') \, V_\varphi^r(t-t';\vecx-\vecx') \, .
\end{align}
These terms introduce a non-locality in time. The term $\pi_\varphi^r(t,t')$ arises due to the condensate self-energy and interactions with $\phi$ and $\chi$ particles, whereas the term $v_\varphi^r(t-t')$ arises from interactions with only $\chi$ particles. In addition, the effective mass is modified.

\paragraph{Multiple-scale analysis for the condensate EoM}
Ref.~\cite{Ai:2023ahr} has analyzed the EoM in Eq.~\eqref{eq:eom-varpphi3} using so-called multiple-scale analysis~\cite{Bender,Holmes} in the case $\pi^r_\varphi(t,t')=\pi^r_\varphi(t-t')$, which applies when $\phi$-particles are in thermal equilibrium and $\Delta_\phi(x,x')=\Delta_\phi(x-x')$.  Assuming that the two terms $\ddot{\varphi}$ and $\overbar{M_\phi}^2\varphi$ are dominant in the EoM, this implies that
\begin{align}
    \overbar{M_\phi}^2\gg \lambda_\phi\varphi^2\,,
\end{align}
 which leads to a quasi-harmonic oscillation regime. As $M^2_\phi \gsim \overbar{M_\phi}^2$ (because the last two terms in Eq.~\eqref{eq:M2} correspond to only the first two diagrams in Eq.~\eqref{eq:Pi-phi}), in this regime one also has
\begin{align}
    M^2_\phi \approx m_\phi^2+\Pi_\phi^\H\,.
\end{align}
That is, the $\varphi$-dependent mass can be neglected.~For practical purposes, one can take the following approximation
\begin{align}
\label{eq:approximation-Mphi}
    M_\phi^2\approx \overbar{M_\phi}^2\approx m_\phi^2+\frac{g}{2}\mathbf{\Delta}_\chi^{++} (0) \overset{{\rm high}\,T}{=\joinrel=} m_\phi^2+\frac{g}{24}T^2\,,
\end{align}
where the term with $\Delta_\phi^{++}$ has been neglected as $\phi$ particles are far from equilibrium in the freeze-in scenario. Below, we neglect the minor difference between $\overbar{M_\phi}$ and $M_\phi$, and therefore do not distinguish $\overbar{M_\phi}$ from $M_\phi$ anymore. Now, we briefly introduce multiple-scale analysis. For more details, the reader is referred to Refs.~\cite{Ai:2021gtg,Wang:2022mvv,Ai:2023ahr}. 

The idea of multiple-scale analysis is to realize that there is a hierarchy in the time scales in the evolution of the condensate. The shorter time scale corresponds to the oscillation frequency, which can also be understood as the microscopic time scale, and the longer time scale corresponds to the damping rate, which can be understood as the macroscopic time scale. The condensate evolution has the form $\varphi(t)=A(t)B(t)$ where $B(t)$ is an oscillating function and $A(t)$ is the envelope function. It is expected that
\begin{align}
   M_\phi \sim \frac{1}{B(t)}\frac{\d B(t)}{\d t} \gg  \frac{1}{A(t)}\frac{\d A(t)}{\d t} \,.
\end{align}
When taking time derivatives of $\varphi$, one can obtain quantities at different orders in magnitude. In order to trace these quantities, one can formally introduce two different time variables, $\varphi(t,\tau)=A(\tau)B(t)$ with $\tau=\varepsilon t$ where $\varepsilon$ is only a bookkeeping parameter. Now we have 
\begin{align}
    \frac{\d\varphi}{\d t}=\frac{\d\tau}{\d t}\frac{\partial\varphi(t,\tau)}{\partial\tau}+\frac{\partial\varphi(t,\tau)}{\partial t}=\varepsilon \frac{\d A(\tau)}{\d\tau} B(t)+A(\tau)\frac{\d B(t)}{\d t}\,.
\end{align}
This way, the smaller term is associated with the bookkeeping factor $\varepsilon$. Meanwhile, one shall assign an $\varepsilon$ for each perturbation term in Eq.~\eqref{eq:eom-varpphi3}. Once the two-time scales have been introduced, we need to replace $\varphi(t)$ in Eq.~\eqref{eq:eom-varpphi3} with $\varphi(t,\tau)$, and $\varphi(t')$ with $\varphi(t',\tau')$. 

To localize the non-local terms in Eq.~\eqref{eq:eom-varpphi3}, one performs a Taylor expansion in the variable $\tau'$ at $\tau'=\tau$. Now, the EoM in Eq.~\eqref{eq:eom-varpphi3} can be solved perturbatively to the leading order in $\varepsilon$. The solution takes the following form:\footnote{If we distinguish $\overbar{M_\phi}$ from $M_\phi$, then in Eq.~\eqref{eq:condensate-form} $M_\phi$ would be replaced by $\overbar{M_\phi}+f(t)$ where $f(t)$ also obeys an equation of motion~\cite{Ai:2023ahr}. The solution for $f(t)$ should be close to $M_\phi-\overbar{M_\phi}$ in the limit of $\lambda_\phi=0$. For $\lambda_\phi\neq 0$, $f(t)$ contains both the additional self-energy corrections and the correction from the term $\lambda_\phi\varphi^3/6$ to the leading-order frequency in multiple-scale analysis, while $M_\phi-\overbar{M_\phi}$ contains only the additional self-energy corrections.}
\begin{align}
\label{eq:condensate-form}
    \varphi(t)\approx A(t) \cos\left[\int ^{t} M_\phi (t') \, \d t'\right] \approx A(t)\cos\left(M_\phi(t) t\right)\,,
\end{align}
where an adiabatic change in $M_\phi(t)$ is assumed. The EoM for $A(t)$ reads~\cite{Ai:2023ahr}
\begin{align}
\label{eq:eom_A(tau)}
\frac{\d A (t)}{\d t}+\left(\gamma +\frac{3}{2}H(t)+\frac{1}{2 M_\phi(t)}\frac{\d M_\phi(t)}{\d t}\right) A(t) +\frac{\sigma}{2} [A(t)]^3&=0\,,
\end{align}
where
\begin{align}
\label{eq:important_quantities}
\gamma \equiv -\frac{\text{Im} [\widetilde{\pi}_\varphi^r (M_\phi)]}{2M_\phi}\,, \qquad \sigma \equiv -\frac{\text{Im} [\widetilde{v}_\varphi^r (2M_\phi)]}{24 M_\phi}\,,
\end{align}
with a tilde indicating the Fourier transform:
\begin{equation}
\label{PiTildeDef}
\widetilde{\pi}_\varphi^r(\omega) = \int_{-\infty}^{\infty} \d t'\, \e^{\i\omega(t-t')} \pi_\varphi^r(t-t') \, ,
\qquad 
\widetilde{v}_\varphi^r(\omega) = \int_{-\infty}^{\infty} \d t' \, \e^{\i\omega(t-t')} v_\varphi^r(t-t') \, .
\end{equation}
Note that $\widetilde{\pi}^r_\varphi(\omega)=\widetilde{\Pi}^r_\varphi(\omega,\veck=0)$ and $\widetilde{v}^r_\varphi(\omega)=\widetilde{V}^r_\varphi(\omega,\veck=0)$. 

In our case, $\pi_\varphi^r(t,t')\neq \pi^r_\varphi(t-t')$, but one can still apply the above multiple-scale analysis. The only difference is that now we have the following replacement
\begin{align}
    \widetilde{\pi}_\varphi^r(\omega)\rightarrow \widetilde{\pi}_\varphi^r(\omega,t) = \int_{-\infty}^{\infty} \d t'\, \e^{\i\omega(t-t')} \pi_\varphi^r(t,t')\,.
\end{align}
and thus $\Tilde{\pi}^r_\varphi(M_\phi)\rightarrow \tilde{\pi}_\varphi^r(M_\phi,t)$ in the definition of $\gamma$ in Eq.~\eqref{PiTildeDef}. Taking the inverse Wigner transform of $\pi_\varphi^r(t,t')$ inside the integral, one obtains 
\begin{align}
    \widetilde{\pi}_\varphi^r(t,\omega) &= \int_{-\infty}^{\infty} \d t'\, \e^{\i\omega(t-t')} \int \frac{\d\omega'}{2\pi} \, \e^{-\i \omega'(t-t')} \overbar{\pi} \left(\omega',\frac{t+t'}{2}\right) \notag \\
    &=\overbar\pi\left(\omega,\frac{t+t'}{2}\right)\notag\\
    &= \overbar{\pi}_\varphi^r(\omega,t)-\frac{(t-t')}{2}\partial_t\overbar{\pi}(\omega,t)+\cdots\,.
\end{align}
Thus, at the leading order of the gradient expansion, $\widetilde{\pi}^r(t,\omega)$ can be identified as the condensate self-energy in Winger space. While for $\Tilde{v}^r_\varphi$ we simply have $\widetilde{v}_\varphi^r(\omega)=\overbar{v}_\varphi^r(\omega)$ due to $v^r_\varphi(t,t')=v^r_\varphi(t-t')$.

The imaginary parts of the retarded self-energy and proper four-vertex function of the condensate, $\gamma$ and $\sigma$ in Eqs.~\eqref{eq:eom_A(tau)} and \eqref{eq:important_quantities}, have an interpretation in terms of microscopic processes. To see this, we note that (see e.g., Ref.~\cite{Anisimov:2008dz})
\begin{align}
\label{eq:ImPi-varphi}
    {\rm Im} \, \overbar{\pi}_\varphi^r= -\frac{\i}{2} \left(\overbar{\pi}_\varphi^{>}-\overbar{\pi}_\varphi^{<}\right)\,,
\end{align}
and similarly for the condensate proper four-vertex function, 
\begin{align}
    {\rm Im} \, \overbar{v}^r_\varphi=-\frac{\i}{2}\left(\overbar{v}^{>}_\varphi-\overbar{v}_\varphi^{<}\right)\,.
\end{align}
This has the same structure as the collision term in Eq.~\eqref{eq:kinetic-eq} except that there are no $\Delta^{<,>}_\phi$. Using the on-shell limit in Eq.~\eqref{eq:on-shell-limit} for $\Delta^{<,>}_\phi$ (with $M_\phi$ given by Eq.~\eqref{eq:approximation-Mphi}) and Eq.~\eqref{eq:Deltachi-varphi0} for $\mathbf{\Delta}_\chi^{<,>}$, one can see that ${\rm Im} \, \overbar{\pi}^{\rm R}_\varphi(t,M_\phi)$ describes all possible process with $\{\varphi \phi\phi\phi\}$ or $\{\varphi\phi \chi\chi\}$ where the condensate quantum has fixed four-momentum $\bar{k}\equiv (M_\phi,\vec{0})$, provided the on-shell conditions can be satisfied (see Appendix~\ref{app:cutting}). For ${\rm Im}\, \overbar{v}_\varphi^{\rm R}(2M_\phi)$, we have the process $\varphi\varphi\leftrightarrow\chi\chi$~\cite{Wang:2022mvv}. 

\section{Boltzmann equations for particles and condensates}
\label{sec:Boltzmann}

\subsection{Recap for the simplified model}

To put the condensate EoM (Eq.~\eqref{eq:eom_A(tau)}) in the form of a Boltzmann equation, note that for the motion~\eqref{eq:condensate-form} if
\begin{align}
   \frac{\dot{A}}{A} \ll M_\phi\,,\quad \frac{\dot{M_\phi}}{M_\phi}\ll M_\phi\,,
\end{align}
which should be valid for most cases of interest,
the energy of the condensate can be approximated as
\begin{align}
\label{eq:rho-varphi}
       \rho _{\varphi}= \frac{1}{2} \dot{\varphi}^2 +\frac{1}{2} M_\phi^2 \varphi^2 \approx \frac{1}{2}M_\phi^2 A(t)^2\,.
\end{align}
Defining $n_\varphi=\rho_\varphi/M_\phi$, we can rewrite Eq.~\eqref{eq:eom_A(tau)} as 
\begin{align}
    \dot{n}_\varphi+3H n_\varphi=-2\gamma n_\varphi-\frac{2\sigma}{M_\phi} n_\varphi^2\,.
\end{align}
Note that the term with $\dot{M_\phi}/M_\phi$ from Eq.~\eqref{eq:eom_A(tau)} is cancelled by the same term from the derivative of $n_\varphi$. Thus, we recover the well-known result for the QCD axion that the number density (not the energy density) is covariantly conserved in the presence of a slowly varying particle mass, once condensate oscillations begin in the regime $M_\phi\gg H$.~We further define the following collision operators 
\begin{align}
    \hat{\C}_\pi=-2\gamma =\frac{{\rm Im}\,\Bar{\pi}^r_\varphi[M_\phi] }{M_\phi}\,,\qquad \hat{\C}_v=-\frac{\sigma}{M_\phi}=\frac{{\rm Im}\,\bar{v}^r_\varphi[2M_\phi]}{24 M^2_\phi}\,.
\end{align}
Then Eq.~\eqref{eq:eom_A(tau)} finally becomes Eq.~\eqref{eq:Boltzman-condensate}. Explicitly, $\hat{\C}_{\pi/v}$ take the form 
\begin{align}
\label{eq:collision-term-condensate}
    \hat{\C}_{\pi/v}[f_i]=&\sum_{\rm all\ possible\ processes}\Bigg[\frac{1}{2M_\phi\times (2M_\phi)^{n-1}}\int \prod_i \frac{\d^3 \vec{k}_i}{(2\pi)^3 \, 2k_i^0} (2\pi)^4 \delta^{(4)}(n\bar{k}+k_{A1}+\cdots-k_{B1}-\cdots)\notag\\
    &\times \{(1\pm f_{A1})\cdots f_{B1}\cdots- f_{A1}\cdots (1\pm f_{B1})\cdots\} \, |\M_{(n\varphi) A_1\cdots\rightarrow B_1\cdots}|^2\Bigg]\,,
\end{align}
where $\Bar{k}=(M_\phi,\vec{0})$ is the four-momentum of condensate quanta, and $n$ counts how many condensate quanta are involved in the interaction. For instance, $n=1$ for $\hat{\C}_\pi$, $n=2$ for $\hat{\C}_v$. At tree-level, $\M_{(n\varphi) A_1\cdots\rightarrow B_1\cdots}$ is read from the Lagrangian by replacing $\varphi$ with $1$. For example, for the term $g\varphi^2\chi^2/4$, we have $|\M_{(2\varphi)\rightarrow \chi\chi}|^2=1/2\times (g/2)^2$ where the particular factor of $1/2$ is due to the exchange symmetry of the produced two $\chi$-particles.

Now let us also formulate the particle Boltzmann equation in Eq.~\eqref{eq:rel-Boltzmann-particle} in terms of the particle density. We initiate our analysis with the first term on the left-hand side, $k_{\mu}\partial^{\mu}_{(x)} f_\phi(k,t)/k^0$.
Since the condensate background is homogeneous, we expect that $f_\phi(k,x)=f_\phi(k^0,t)$. After imposing the on-shell condition, we simply have
\begin{equation}
\label{1lefthandside}
    \frac{1}{k^0}k_{\mu}\partial^{\mu}_{(x)} f_\phi(k,t)\rightarrow \frac{\partial f_\phi(E,t)}{\partial t}\,,
\end{equation}
where $E=\sqrt{M_\phi^2+\veck^2}$. One can also view $f_\phi$ as a function of $|\veck|$, $f_\phi(|\veck|,t)$.
So far, we have not taken into account the expansion of the Universe in this equation. As commonly done in the literature, we only take into account the effect of the Universe expansion in the kinetic term (the first term) of Eq.~\eqref{eq:rel-Boltzmann-particle} while the collision terms are computed in flat spacetime. This artificial combination is partly due to difficulties in deriving the dynamics in a curved spacetime background from the beginning but also is arguably valid since the scale of microscopic processes is much smaller than the Hubble scale and the impact of the expansion of the Universe can be neglected. To do that, we do the following replacement~\cite{Dodelson:2003ft}
\begin{equation}
\label{eq:89}
   \frac{\partial f_{\phi}(|\vec{k}|,t)}{\partial t}\rightarrow \frac{\partial f_\phi(|\veck|,t)}{\partial t} - H |\veck| \frac{\partial f_\phi(|\veck|,t)}{\partial |\veck|}\,.
\end{equation}
After invoking the definition 
\begin{align}
  n_\phi(t)=\int \frac{\d^3\veck}{(2\pi)^3} \, f_\phi(|\veck|,t)\,, 
\end{align} 
and integrating Eq.~\eqref{eq:89} yields
\begin{equation}
\label{eq:91}
    \dot{n}_{\phi}(t) +3H n_{\phi}-\int\frac{\d^3\veck}{(2\pi)^3}\left(\frac{1}{2}\frac{1}{E} \frac{\d M_\phi^2}{\d t}\frac{\partial f_\phi(E,t)}{\partial E}\right)\,,
\end{equation}
where the last term is due to $M_\phi$ being time dependent. This term will be cancelled by the second term in Eq.~\eqref{eq:rel-Boltzmann-particle} that we discuss below.
We assume that the expansion of the Universe does not significantly modify the second and the collision term in Eq.~\eqref{eq:rel-Boltzmann-particle}.

The second term on the left-hand side of Eq.~\eqref{eq:rel-Boltzmann-particle} is usually identified as a force term by writing 
\begin{align}
    \frac{1}{2}(\partial^{(x)}_\mu M_\phi^2)=M_\phi F_\mu\,, \quad F_\mu\equiv\frac{\partial M_\phi}{\partial x^\mu}\,.
\end{align}
This force is due to the spacetime dependence of the particle mass under study. In our case, due to the assumption of a homogeneous background, we simply have 
\begin{equation}
\label{eq:93}
 \frac{1}{2}\frac{1}{E} \frac{\d M_\phi^2}{\d t}\frac{\partial f_\phi(E,t)}{\partial E}\,.
\end{equation}
Doing the integral over $\d^3\veck$, we get a term that is exactly cancelled by the last term in Eq.~\eqref{eq:91}. One then finally get Eq.~\eqref{eq:Boltz-number-density}. Generally, the right-hand side can only be discussed on a case-by-case basis. Comparing this equation with Eq.~\eqref{eq:Boltzman-condensate}, we see that in deriving the latter, there is no need to do an integral over the momentum space. This is reasonable as the condensate is cold and all condensate quanta have a fixed, unique four-momentum $\Bar{k}=(M_\phi,\vec{0})$. However, if one wishes, he/she may note that condensate quanta have a singular distribution function given by 
\begin{align}
    f_\varphi(\veck, t)= n_\varphi(t) (2\pi)^3\delta^{(3)}(\veck)\,,
\end{align}
which cannot be obtained from the Bose-Einstein distribution in the zero-temperature limit.

Until now, we have not been concerned with the collision terms  that contain explicit factors of $\varphi$ in the particle Boltzmann equation. We analyze this in the next subsection.

\subsection{Collision terms in the presence of condensates}
\label{sec:collision-with-condensate}

Now we consider the $\phi$ particle self-energies involving the condensate, i.e., the third and fourth diagrams in Eq.~\eqref{eq:Pi-phi}, or equivalently, the third and fourth terms in Eq.~\eqref{eq:Pi-phi-ex}. We consider the quasi-harmonic oscillation regime in which the background field can be written as (c.f. Eq.~\eqref{eq:condensate-form})
\begin{align}
    \varphi(t)=\frac{A(t)}{2} \left(\e^{\i \bar{k} x}+\e^{-\i\bar{k}x}\right)\,.
\end{align}
Recall that $\Bar{k}=(M_\phi,\vec{0})$.
To be specific, let us consider the fourth term in Eq.~\eqref{eq:Pi-phi-ex}, namely
\begin{align}
    \Pi_{\phi,(4)}^{ab}(x_1,x_2)=-\frac{\i g^2}{2} \varphi(x_1)[\Delta_\chi^{ab}(x_1,x_2)]^2\varphi(x_2)\equiv \varphi(x_1) \mathbf{\Pi}^{ab}_{\phi,(4)}(x_1,x_2)\varphi(x_2)\,,
\end{align}
where we have defined $\mathbf{\Pi}^{ab}_{\phi,(4)}$; we refer to it as the {\it reduced self-energy}. We have suppressed the Keldysh index on $\varphi$ as in the EoMs, one would take $\varphi^+=\varphi^-=\varphi$ in the end. One can also define the reduced self-energy for the third term in Eq.~\eqref{eq:Pi-phi-ex}, i.e., $\mathbf{\Pi}_{\phi,(3)}^{ab}$.

When performing the Wigner transform for $\Pi^{<,>}_{\phi,(4)}(x_1,x_2)$, we have a factor in the integral that looks like
\begin{align}
    \varphi\left(x+\frac{r}{2}\right)\varphi\left(x-\frac{r}{2}\right)=\frac{A^2(t)+\O(\partial^2)}{4}\left[\cos(2M_\phi t)+\e^{\i\bar{k}r}+\e^{-\i\bar{k}r}\right]\,.
\end{align}
The self-energy in Wigner space reads
\begin{align}
\label{eq:Pi-4-Wigner}
    \overbar{\Pi}_{\phi,(4)}^{<,>}(k,x) &=\frac{A^2(t)\cos(2M_\phi t)}{4} \overbar{\mathbf{\Pi}}_{\phi,(4)}^{<,>}(k,x) + \frac{A^2(t)}{4} \overbar{\mathbf{\Pi}}_{\phi,(4)}^{<,>}(k+\bar{k},x) \nonumber \\
    &\quad +\frac{A^2(t)}{4} \overbar{\mathbf{\Pi}}_{\phi,(4)}^{<,>}(k-\bar{k},x)\,.
\end{align}
Substituting the above into the right-hand side of Eq.~\eqref{eq:kinetic-eq} in the on-shell limit, one then obtains three different types of collision terms:  
\begin{itemize}
    \item[(a)] {\it Condensate modified vertex}: The first term in Eq.~\eqref{eq:Pi-4-Wigner} gives collision terms with two $\chi$ particles and one $\phi$ particle. The squared transition amplitude is multiplied by $A^2(t)\cos(2M_\phi t)/4$ compared with that when $\varphi(t)=1$. If one takes the cycle average, this collision term vanishes.
    \item[(b)] {\it Absorbing one condensate quantum}: The second term describes a condensate quantum absorbed by particles. The squared transition amplitude is enhanced by a factor of $A^2(t)/4$ compared to that when $\varphi(t)=1$. We denote this type of collision term as $\C_{(+\varphi)}[f_\phi,f_i]$. If we define
    \begin{align}
\label{eq:collision-term-condensate2}
    &\hat{\C}_{\phi\varphi A_1\cdots\rightarrow B_1\cdots }[f_\phi,f_i]=\sum\Bigg[\frac{1}{2k^0\times (2M_\phi)}\int \prod_i \frac{\d^3 k_i}{(2\pi)^3 \, 2k_i^0} (2\pi)^4 \delta^{(4)}(k+\bar{k}+k_{A1}+\cdots-k_{B1}-\cdots)\notag\\
    &\times \{(1\pm f_\phi)(1\pm f_{A1})\cdots f_{B1}\cdots-f_\phi f_{A1}\cdots (1\pm f_{B1})\cdots\} \, |\M_{\phi\varphi A_1\cdots\rightarrow B_1\cdots}|^2\Bigg]\,,
\end{align}
where $|\M_{\phi\varphi...\rightarrow ...}|$ is defined as described below Eq.~\eqref{eq:collision-term-condensate} such that
    \begin{align}
        \C_{(+\varphi)}[f_\phi,f_i]=n_\varphi \hat{\C}_{\phi\varphi A_1\cdots\rightarrow B_1\cdots}\,.
    \end{align}
\item[(c)] {\it Emitting one condensate quantum}: Similar to the last case, the third term in Eq.~\eqref{eq:Pi-4-Wigner} describes a condensate quantum emitted by particles. We denote this type of collision term as $\C_{(-\varphi)}[f_\phi,f_i]$. Similarly, one has
\begin{align}
    \C_{(-\varphi)}[f_\phi,f_i]=n_\varphi\hat{\C}_{\phi A_1\cdots\rightarrow \varphi B_1\cdots}\,.
\end{align}
\end{itemize}
In all the above, we observe Bose enhancement of the collision term by the condensate number density, $n_\varphi$.

\subsection{Generalisation: the Primakoff interaction}\label{sec:primakoff}

Although the preceding analysis is based on the simple model in Eq.~\eqref{eq:model}, the results are general. Here, we apply our results to an axion-like model. Consider the Lagrangian
\begin{align}
\label{eq:model2}
    \L = \frac{1}{2} \partial_\mu \Phi \partial^\mu \Phi-\frac{1}{2}m^2 \Phi^2 -\frac{1}{4!}\lambda \Phi^4-\frac{g}{4} \Phi F_{\mu\nu}\widetilde{F}^{\mu\nu}\,,
\end{align}
where $F_{\mu\nu}=\partial_\mu A_\nu-\partial_\nu A_\mu$ and $\widetilde{F}$ is its Hodge dual.
We assume that the axion field $\Phi$ forms a condensate, $\langle \Phi\rangle\equiv \varphi$, whose oscillation is quasi-harmonic.  Substituting $\Phi=\varphi+\phi$ into the Lagrangian, one obtains
$\L=\bar{\L}[\varphi]+\hat{\L}[\phi;\varphi]$ where 
\begin{align}
    \bar{\L}[\varphi]=\frac{1}{2}(\partial_\mu \varphi)\partial^\mu \varphi-\frac{1}{2}m^2 \varphi^2 -\frac{1}{4!}\lambda \varphi^4\,,
\end{align}
and 
\begin{align}
\label{eq:L-hatphi}
    \hat{\L}[\phi;\varphi]&=\frac{1}{2}(\partial_\mu \phi)\partial^\mu \phi-\frac{1}{2}\left(m^2+\frac{\lambda}{2}\varphi^2\right) \phi^2-\frac{\lambda}{3!}\varphi\phi^3-\frac{1}{4!}\lambda\phi^4-\frac{g}{4} \phi F_{\mu\nu}\widetilde{F}^{\mu\nu}-\frac{g}{4} \varphi F_{\mu\nu}\widetilde{F}^{\mu\nu}\notag\\
    &\hspace{5mm} +({\rm linear\ terms\ in\ fluctuations})\,.
\end{align}
We also need the interaction between photons and electrons 
\begin{align}
    \L_{\rm SM}\supset \bar{\psi}(\i \slashed{\partial}-m+e\slashed{A})\psi \,.
\end{align}

Properly truncating the effective action, we have (only diagrams with at least one condensate quantum or $\phi$ propagator are shown)
\begin{align}
\label{eq:Gamma2-axion-model}
\Gamma_2=-\i\left(\,
\begin{tikzpicture}[baseline={-0.025cm*height("$=$")}]
\filldraw (0,0) circle (1.5pt) node {} ;
\draw[draw=black,dashed] (-0.5,0) circle (0.5);
\draw[draw=black,dashed] (0.5,0) circle (0.5);
\end{tikzpicture}
\, +\,
\begin{tikzpicture}[baseline={-0.025cm*height("$=$")}]
\draw[draw=black,thick] (-0.5,0) circle (0.15) ;
\draw[draw=black,thick] (-0.5-0.12,0+0.12) -- (-0.5+0.12,0-0.12);
\draw[draw=black,thick] (-0.5-0.12,0-0.12) -- (-0.5+0.12,0+0.12);
\draw[draw=black,dashed] (-0.35,0) -- (0.25,0);
\filldraw (0.25,0) circle (1.5pt) node {} ;
\filldraw (1.25,0) circle (1.5pt) node {} ;
\draw[draw=black,dashed] (0.75,0) circle (0.5);
\draw[dashed] (0.25,0) -- (1.25,0);
\draw[draw=black,dashed] (1.25,0) -- (1.85,0);
\draw[draw=black,thick] (2,0) circle (0.15) ;
\draw[draw=black,thick] (2-0.12,0+0.12) -- (2+0.12,0-0.12);
\draw[draw=black,thick] (2-0.12,0-0.12) -- (2+0.12,0+0.12);
\end{tikzpicture}
\, +\, 
\begin{tikzpicture}[baseline={-0.025cm*height("$=$")}]
\draw[double,decorate,decoration={snake,amplitude=.5mm,segment length=1.5mm}] (0.75,0) circle (0.5);
\draw[draw=black,dashed] (0.25,0) -- (1.25,0); 
\filldraw (0.25,0) circle (1.5pt) node {} ;
\filldraw (1.25,0) circle (1.5pt) node {} ;
\end{tikzpicture}
\, +\,
\begin{tikzpicture}[baseline={-0.025cm*height("$=$")}]
\filldraw (0.25,0) circle (1.5pt) node {} ;
\filldraw (1.25,0) circle (1.5pt) node {} ;
\draw[double,decorate,decoration={snake,amplitude=.5mm,segment length=1.5mm}] (0.25,0) arc (180:120:0.5);
\draw[double,decorate,decoration={snake,amplitude=.5mm,segment length=1.5mm}] (1.25,0) arc (0:60:0.5);
\draw[draw=black,thick] (0.75,0.45) circle (0.25);
\filldraw (1,0.45) circle (1.5pt) node {} ;
\filldraw (0.5,0.45) circle (1.5pt) node {} ;
\draw[double,decorate,decoration={snake,amplitude=.5mm,segment length=1.5mm}] (0.25,0) arc (-180:0:0.5);
\draw[draw=black,dashed] (0.25,0) -- (1.25,0);
\end{tikzpicture}\right)
\,,
\end{align}
where a dashed/solid line denotes the $\phi$/electron propagator.
Above, we have used the double snaked line
to denote the $\varphi$-dependent photon propagator. If we do the small-field expansion, we would get the following diagram from the photon {\it one-loop} diagram,
\begin{align}
\label{eq:condensate-decay-to-photons}
\begin{tikzpicture}[baseline={-0.025cm*height("$=$")}]
\filldraw (0.25,0) circle (1.5pt) node {} ;
\filldraw (1.25,0) circle (1.5pt) node {} ;
\draw[draw=black,thick] (-0.5,0) circle (0.15) ;
\draw[draw=black,thick] (-0.5-0.12,0+0.12) -- (-0.5+0.12,0-0.12);
\draw[draw=black,thick] (-0.5-0.12,0-0.12) -- (-0.5+0.12,0+0.12);
\draw[draw=black,dashed] (-0.35,0) -- (0.25,0);
\draw[decorate,decoration={snake,amplitude=.5mm,segment length=1.5mm}] (0.75,0) circle (0.5);
\draw[draw=black,dashed] (1.25,0) -- (1.85,0);
\draw[draw=black,thick] (2,0) circle (0.15) ;
\draw[draw=black,thick] (2-0.12,0+0.12) -- (2+0.12,0-0.12);
\draw[draw=black,thick] (2-0.12,0-0.12) -- (2+0.12,0+0.12);
\end{tikzpicture}\,,
\end{align}
where the single snaked line denotes the leading order of the photon propagator after the small-field expansion.
This diagram gives the decay of the axion condensate into photons. A diagram similar to the last one in Eq.~\eqref{eq:Gamma2-axion-model} but without the $\phi$ propagator will generate the following diagram in the small-field expansion,
\begin{align}
\label{eq:condensate-Primarkoff}
\begin{tikzpicture}[baseline={-0.025cm*height("$=$")}]
\filldraw (0.25,0) circle (1.5pt) node {} ;
\filldraw (1.25,0) circle (1.5pt) node {} ;
\draw[draw=black,thick] (-0.5,0) circle (0.15) ;
\draw[draw=black,thick] (-0.5-0.12,0+0.12) -- (-0.5+0.12,0-0.12);
\draw[draw=black,thick] (-0.5-0.12,0-0.12) -- (-0.5+0.12,0+0.12);
\draw[draw=black,dashed] (-0.35,0) -- (0.25,0);
\draw[decorate,decoration={snake,amplitude=.5mm,segment length=1.5mm}] (0.25,0) arc (180:120:0.5);
\draw[decorate,decoration={snake,amplitude=.5mm,segment length=1.5mm}] (1.25,0) arc (0:60:0.5);
\draw[draw=black,thick] (0.75,0.45) circle (0.25);
\filldraw (1,0.45) circle (1.5pt) node {} ;
\filldraw (0.5,0.45) circle (1.5pt) node {} ;
\draw[decorate,decoration={snake,amplitude=.5mm,segment length=1.5mm}] (0.25,0) arc (-180:0:0.5);
\draw[draw=black,thick] (2,0) circle (0.15) ;
\draw[draw=black,dashed] (1.25,0) -- (1.85,0);
\draw[draw=black,thick] (2-0.12,0+0.12) -- (2+0.12,0-0.12);
\draw[draw=black,thick] (2-0.12,0-0.12) -- (2+0.12,0+0.12);
\end{tikzpicture}\,.
\end{align}
For the third and fourth diagrams in Eq.~\eqref{eq:Gamma2-axion-model}, we simply replaced the double snaked lines in Eq.~\eqref{eq:Gamma2-axion-model} with single snaked lines after the small-field expansion.

\paragraph{Condensate Boltzmann equation}

For the collision terms in the condensate Boltzmann equation in Eq.~\eqref{eq:Boltzman-condensate}, we simply cut the second diagram in Eq.~\eqref{eq:Gamma2-axion-model} and the diagrams in Eqs.~\eqref{eq:condensate-decay-to-photons} and~\eqref{eq:condensate-Primarkoff}, see Appendix~\ref{app:cutting}. For the first two diagrams, the kinematically allowed processes are the pure decay $\varphi\rightarrow \gamma\gamma$ and scattering process $\varphi \phi\rightarrow \phi\phi$, and their inverse processes. They belong to the $\hat{\C}_\pi$ type interactions since only one condensate quantum is involved. One can write down the collision terms following Eq.~\eqref{eq:collision-term-condensate}. The only difference one needs to take care of is the polarization of the photons and the spinor structure of electrons in writing down $\M$. This complication also appears for conventional Boltzmann equations in the absence of condensates and we do not discuss it further. After cutting the diagram in Eq.~\eqref{eq:condensate-Primarkoff}, we obtain the Primakoff process with the axion {\it particle} replaced by an axion condensate {\it quantum}, $\gamma e\rightarrow \varphi e$ and its inverse process. Of course, one also has the scattering $\varphi \gamma \leftrightarrow \bar{e} e$ and the decay process $\varphi\leftrightarrow \bar{e}e \gamma$ (if kinematically allowed).

Through the above processes, condensate quanta can be created and annihilated. One may naively conclude that the Primakoff interaction, for example, could create an equilibrium number density of both particles \emph{and} condensate. However, generically, it is expected that the total effect will lead to a decrease in the total condensate quantum number, leading to the dissipation of the condensate energy into the plasma.  This expectation is based on the fact that the cold condensate has zero entropy and therefore, thermodynamically, dissipation and thermalization occur instead of the reverse. This is also explicitly confirmed in the studies of the evolution of oscillating condensates using the model in Eq.~\eqref{eq:model}~\cite{Ai:2021gtg,Wang:2022mvv,Ai:2023ahr}, as well as in studies on reheating.

\paragraph{Particle Boltzmann equation}

For the collision terms in the particle Boltzmann equation in Eq.~\eqref{eq:rel-Boltzmann-particle}, we look at the last three diagrams in Eq.~\eqref{eq:Gamma2-axion-model} which gives
\begin{align}
\label{eq:Pi-phi-axion-model}
\Pi_\phi\supset \i \Bigg(\,
\begin{tikzpicture}[baseline={-0.025cm*height("$=$")}]
\draw[draw=black,dashed] (-0.35,0) -- (0.25,0);
\draw[draw=black,thick] (0.25,-0.6) circle (0.15) ;
\draw[draw=black,thick] (-0.5-0.12+0.75,0+0.12-0.6) -- (-0.5+0.12+0.75,0-0.12-0.6);
\draw[draw=black,thick] (-0.5-0.12+0.75,0-0.12-0.6) -- (-0.5+0.12+0.75,0+0.12-0.6);
\draw[draw=black,dashed] (0.25,0) -- (0.25,-0.45);
\draw[draw=black,dashed] (0.75,0) circle (0.5);
\filldraw (0.25,0) circle (1.5pt) node {} ;
\filldraw (1.25,0) circle (1.5pt) node {} ;
\draw[draw=black,thick] (1.25,-0.6) circle (0.15) ;
\draw[draw=black,thick] (-0.5-0.12+0.75+1,0+0.12-0.6) -- (-0.5+0.12+0.75+1,0-0.12-0.6);
\draw[draw=black,thick] (-0.5-0.12+0.75+1,0-0.12-0.6) -- (-0.5+0.12+0.75+1,0+0.12-0.6);
\draw[draw=black,dashed] (1.25,0) -- (1.85,0);
\draw[draw=black,dashed] (1.25,0) -- (1.25,-0.45);
\end{tikzpicture}
\,+\,
\begin{tikzpicture}[baseline={-0.025cm*height("$=$")}]
\draw[draw=black,dashed] (-0.35,0) -- (0.25,0);
\draw[decorate,decoration={snake,amplitude=.5mm,segment length=1.5mm}] (0.75,0) circle (0.5);
\filldraw (0.25,0) circle (1.5pt) node {} ;
\filldraw (1.25,0) circle (1.5pt) node {} ;
\draw[draw=black,dashed] (1.25,0) -- (1.85,0);
\end{tikzpicture}
\, +\,
\begin{tikzpicture}[baseline={-0.025cm*height("$=$")}]
\filldraw (0.25,0) circle (1.5pt) node {} ;
\filldraw (1.25,0) circle (1.5pt) node {} ;
\draw[decorate,decoration={snake,amplitude=.5mm,segment length=1.5mm}] (0.25,0) arc (180:120:0.5);
\draw[decorate,decoration={snake,amplitude=.5mm,segment length=1.5mm}] (1.25,0) arc (0:60:0.5);
\draw[draw=black,thick] (0.75,0.45) circle (0.25);
\filldraw (1,0.45) circle (1.5pt) node {} ;
\filldraw (0.5,0.45) circle (1.5pt) node {} ;
\draw[decorate,decoration={snake,amplitude=.5mm,segment length=1.5mm}] (0.25,0) arc (-180:0:0.5);
\draw[dashed] (0.25,0) -- (-0.35,0);
\draw[dashed] (1.25,0) -- (1.85,0);
\end{tikzpicture}
\Bigg) 
\,,
\end{align}
where we have kept the external legs, which do not contribute to the explicit expression of the self-energy, for the reason of applying the cutting rules (see Appendix~\ref{app:cutting}). Cutting the last two diagrams gives the decay process $\phi\leftrightarrow \gamma\gamma$, $\phi\leftrightarrow \gamma \Bar{e}e$ (if kinematically allowed), and scattering $\phi\gamma\leftrightarrow \bar{e}e$, $\gamma e \leftrightarrow \phi e$ (the Primarkoff process). The process corresponding to the first diagram has been discussed in the two-scalar model in Eq.~\eqref{eq:model}.

\section{Summary and Conclusions}
\label{sec:Conc}

In this paper, we have derived from first principles of non-equilibrium quantum field theory the coupled equations describing the interaction between particles and an oscillating homogeneous condensate for a scalar field $\Phi$ in the plasma of the early Universe. This formalism can be used to describe the production of scalar DM, including axions, in the form of both particles, for which an irreducible population is produced by freeze-in through portal interactions with the Standard Model, and of a condensate, which arises due to spontaneous symmetry breaking and/or inflationary fluctuations. The condensate and particles are described by the one- and two-point functions, respectively.

The fundamental non-local EoMs, which in principle were known prior to our work, are, however, very difficult to apply in practice to address questions about the DM relic density. This is due to the equations being non-local (and thus insusceptible to standard numerical solution methods), and due to the technically difficult formalism, which may obscure the essential physics relevant for DM. The derivation of the Boltzmann equation in the absence of condensates is known in the literature~\cite{Calzetta:1986cq,Prokopec:2003pj,Prokopec:2004ic,Drewes:2012qw,Sheng:2021kfc}. However, including the condensate in this derivation is, to the best of our knowledge, lacking.  We thus presented a novel simultaneous Markovianization of the quantum equations for both the one- and two-point functions. The gradient expansion is not sufficient to localize the EoM of the oscillating condensate and a key additional input is the use of multiple-scale analysis introduced in Refs.~\cite{Ai:2021gtg,Wang:2022mvv,Ai:2023ahr} to solve non-equilibrium QFT problems. This Markovianization process, surprisingly, results in coupled Boltzmann equations for {\it both} the condensate and particles with explicit local collision operators. Thus, they can be easily applied to studying the DM relic density.

Studies on the evolution of axion condensates using the CTP formalism have been carried out in Refs.~\cite{Cao:2022bua,Cao:2022kjn} recently. Similarly to Refs.~\cite{Ai:2021gtg,Wang:2022mvv,Ai:2023ahr}, these works assume that the condensate self-energies $\Pi_\varphi$ depend on two-point functions of thermal bath fields only, i.e., fields that are in thermal equilibrium. This way, the condensate EoM can be solved independently. Since $\Pi_\varphi$ depends on $\Delta_\phi$ also, in that computation one thus has to assume a thermal equilibrium form for $\Delta_\phi$. Equivalently,  the (axion) DM {\it particles} are assumed to be in thermal equilibrium with the plasma. Therefore, those results cannot be applied to the freeze-in scenario of DM production. The latter is precisely the motivation of the present work. Hence, in the present work we have to solve the coupled EoMs for the one- and two-point functions simultaneously, i.e. making no assumption on the form of $\Delta_\phi$. We also note that recently the CTP formalism has been used to derive the transport equation of photons in Magnetised Plasmas with axion sources~\cite{McDonald:2023ohd}.

The highly non-trivial non-locality of the quantum EoMs is expressed by the infinite number of derivatives needed to describe them in the gradient expansion. This can be contrasted to a more trivial form of non-locality in the classical equations for a self-gravitating condensate. The condensate ``wavefunction'', $\psi(x,t)$, obeys a non-local (in space) Schr\"{o}dinger equation, where the potential is given by the long-range Newtonian self-gravity of the condensate~\cite{Guth:2014hsa}. The non-locality can be trivially accounted for by introducing a local second-order auxiliary equation for the gravitational potential: Poisson's equation. However, the Newtonian potential introduces no new dynamical degrees of freedom. The coupled equations are local and the system can then be solved by leap-frog spectral methods or finite differences at finite order in derivatives (e.g., Refs.~\cite{Schive:2014dra,Edwards:2018ccc,Chen:2020cef}). In this treatment, the fluctuations upon the condensate have not been accounted for, which makes the condensate EoM local.

Such a simplification cannot be obtained for the full quantum non-locality in QFT. For example, a QFT system is in principle described by a {\it wavefunctional} instead of a wavefunction. Specifically, to fully capture all the information in a QFT, one in principle needs all $n$-point functions which are non-local for $n>1$, although, for most practical purposes, we only need to study the EoMs for the one- and two-point functions. Still, their EoMs  involve non-local integrals involving two-point functions resulting from interactions among particles, or between particles and the condensate, that are characterized by various self-energies. As well as being non-local in time, the quantum equations with infinite derivatives also have a problem concerning the Cauchy data. The process of Markovianization we developed here allows these problems to be bypassed in a controlled manner.\footnote{We thank Dima Levkov for the discussion on these points.}

Our results may find applications in DM production during perturbative reheating~\cite{Chung:1998rq,Giudice:2000ex,Moroi:2020bkq,Garcia:2020wiy,Becker:2023tvd}, although particle production, in that case, may be dominated by the earlier nonperturbative preheating stage~\cite{Dolgov:1989us,Traschen:1990sw,Kofman:1994rk,Shtanov:1994ce,Boyanovsky:1996sq,Kofman:1997yn,Lebedev:2021xey,Lebedev:2021tas}. We emphasize that our derivation assumes that the condensate oscillates quasi-harmonically. For more general oscillations, a first-principles derivation of the dynamics would require further developments in multiple-scale analysis applied to solving the non-local condensate EoM. For DM production in situations other than reheating, however, quasi-harmonically condensates should be a generic enough assumption.

In this work, we have presented only the formalism to describe the interaction between particles and condensates during DM production. In a follow-up paper, we will numerically solve the coupled equations. The condensate can transfer energy to the particles, and condensate quanta are also created in particle processes. In applications with multiple-component DM, these effects could be significant, especially in systems involving resonances and level-crossing, such as those explored in relation to the condensate dynamics alone in Refs.~\cite{Cyncynates:2021xzw,Cyncynates:2023esj,Murai:2023xjn}.

\section*{Acknowledgments}

The work of W.\,Y.\,A. is supported by the UK Engineering and Physical Sciences Research Council (EPSRC), under Research Grant No.~EP/V002821/1.~A.\,B. is supported by the Science and Technology Facilities Council (STFC) Grant No.~ST/T00679X/1.~A.\,M. is supported by an STFC studentship. D.\,J.\,E.\,M. is supported by an Ernest Rutherford Fellowship from the STFC, Grant No.~ST/T004037/1 and by a Leverhulme Trust Research Project (RPG-2022-145).

\appendix

\section{Feynman rules}
\label{app:Feynman}

In this appendix, we specify the Feynman rules for the Lagrangian given in Eq.~\eqref{eq:Sphichi}.
Due to the presence of the condensate, it is more convenient to use the Feynman rules in position space. The propagators are denoted as ($a,b=\pm$)
\begin{equation}
\Delta_\phi^{ab}(x_1,x_2) =
\begin{tikzpicture}[baseline={0cm-0.5*height("$=$")}]
\draw[thick] (-0.6,0) -- (0.6,0) ;
\filldraw (-0.6,0) circle (1.5pt) node[below] {$\scriptstyle(x_1,a)$} ;
\filldraw (0.6,0) circle (1.5pt) node[below] {$\scriptstyle(x_2,b)$} ;
\end{tikzpicture}
\ ,
\qquad\quad
\Delta_\chi^{ab} (x_1,x_2) =
\begin{tikzpicture}[baseline={0cm-0.5*height("$=$")}]
\draw[thick,double,dashed] (-0.6,0) -- (0.6,0) ;
\filldraw (-0.6,0) circle (1.5pt) node[below] {$\scriptstyle(x_1,a)$} ;
\filldraw (0.6,0) circle (1.5pt) node[below] {$\scriptstyle(x_2,b)$} ;
\end{tikzpicture}
\,,
\end{equation}
and the vertices are denoted as,
\begin{align}
&
\begin{tikzpicture}[baseline={-0.025cm*height("$=$")}]
\draw[draw=black,thick] (-0.75, 0.75) -- (0.75,-0.75);
\draw[draw=black,thick] (-0.75,-0.75) -- (0.75,0.75);
\filldraw (0,0) circle (1.5pt) node[below] {$\scriptstyle a$} ;
\end{tikzpicture}
=
a (-i)\lambda_\phi \int \d^{4}x \,,\qquad 
\begin{tikzpicture}[baseline={-0.025cm*height("$=$")}]
\draw[draw=black,double,dashed] (-0.75, 0.75) -- (0.75,-0.75);
\draw[draw=black,double,dashed] (-0.75,-0.75) -- (0.75,0.75);
\filldraw (0,0) circle (1.5pt) node[below] {$\scriptstyle a$} ;
\end{tikzpicture}
=
a (-i)\lambda_\chi \int \d^{4}x \,, \quad
\begin{tikzpicture}[baseline={-0.025cm*height("$=$")}]
\draw[draw=black,thick] (-0.75, 0.75) -- (0.75,-0.75);
\draw[draw=black,double,dashed] (-0.75,-0.75) -- (0.75,0.75);
\filldraw (0,0) circle (1.5pt) node[below] {$\scriptstyle a$} ;
\end{tikzpicture}
=
a (-i)g \int \d^{4}x \,, \\
&
\begin{tikzpicture}[baseline={-0.025cm*height("$=$")}]
\draw[draw=black,thick] (-0.5,0) circle (0.15) ;
\draw[draw=black,thick] (-0.5-0.12,0+0.12) -- (-0.5+0.12,0-0.12);
\draw[draw=black,thick] (-0.5-0.12,0-0.12) -- (-0.5+0.12,0+0.12);
\draw[draw=black,thick] (-0.35,0) -- (0.25,0);
\draw[draw=black, thick] (0.25,0) -- (1,0.85);
\draw[draw=black, thick] (0.25,0) -- (1,-0.85);
\draw[draw=black,thick] (0.25,0) -- (1.3,0);
\filldraw (0.25,0) circle (1.5pt) node[below] {$\scriptstyle a$};
\end{tikzpicture}
=a (-\i) \lambda_\phi \int\d^4 x\, \varphi^a(x)\,,\quad
\begin{tikzpicture}[baseline={-0.025cm*height("$=$")}]
\draw[draw=black,thick] (-0.5,0) circle (0.15) ;
\draw[draw=black,thick] (-0.5-0.12,0+0.12) -- (-0.5+0.12,0-0.12);
\draw[draw=black,thick] (-0.5-0.12,0-0.12) -- (-0.5+0.12,0+0.12);
\draw[draw=black,thick] (-0.35,0) -- (0.25,0);
\draw[draw=black,double,dashed] (0.25,0) -- (1,0.85);
\draw[draw=black,double,dashed] (0.25,0) -- (1,-0.85);
\draw[draw=black,thick] (0.25,0) -- (1.3,0);
\filldraw (0.25,0) circle (1.5pt) node[below] {$\scriptstyle a$};
\end{tikzpicture}
=a (-\i) g \int\d^4 x\, \varphi^a(x)\,.
\end{align}
Above, the use of a double dashed line is to emphasize that the $\chi$ propagator implicitly depends on $\varphi$ due to the modified dispersion relation in Eq.~\eqref{eq:G-inverse-A}. In Sec.~\ref{sec:small-field}, we discuss the expansion of $\Delta_\chi[\varphi]$ in $\varphi$, the small-field expansion from Ref.~\cite{Ai:2021gtg}, and we use the single dashed line for $\mathbf{\Delta}_\chi\equiv \Delta_\chi[\varphi=0]$. For $\Delta_\phi$, we do not need to do such an expansion since we solve the EoM of $\Delta_\phi$ anyway.

A diagram is called one-particle-reducible if it can be separated into two disconnected parts by cutting one propagator. Such diagrams are generated by interaction terms linear in the fluctuation fields. An example is shown in Eq.~\eqref{eq:tadpole}. A diagram is called two-particle-reducible if it can be separated into two disconnected parts by cutting two propagators. An example is the following,
\begin{align}
    \begin{tikzpicture}[baseline={-0.025cm*height("$=$")}]
\draw[draw=black,thick] (-0.5,0) circle (0.5);
\filldraw (0,0) circle (1.5pt) node {} ;
\draw[draw=black,thick] (0.5,0) circle (0.5);
\filldraw (1,0) circle (1.5pt) node {} ;
\draw[draw=black,thick] (1.5,0) circle (0.5) ; 
\end{tikzpicture}
\,.
\end{align}
It is separated into two disconnected parts if one cuts the two propagators of the middle loop. In 2PI effective actions, one only considers two-particle-{\it irreducible} diagrams in $\Gamma_2$ which represents the contribution to the effective action $\Gamma_{\rm 2PI}$ starting from two loops.

\section{Cutting rules}
\label{app:cutting}

In this appendix, we illustrate the cutting rules in more detail using some examples. 

\subsection{Particle kinetic equation}

We first consider how to get the collision term in Eq.~\eqref{eq:rel-Boltzmann-particle} from Eq.~\eqref{eq:kinetic-eq}. Although this is a simple mathematical exercise by simply substituting the on-shell Wightman functions in Wigner space into Eq.~\eqref{eq:kinetic-eq}. It is nonetheless helpful to write down the collision terms based on some rules that can be illustrated with Feynman diagrams.

\paragraph{Without condensate} First, we start with a diagram in $\Gamma_2$, for e.g., the sixth diagram in Eq.~\eqref{eq:Gamma2} (take the leading order after the small-field expansion in $\Delta_\chi$),
\begin{align}
    \begin{tikzpicture}[baseline={-0.025cm*height("$=$")}]
\draw[thick] (0.75,0) circle (0.5);
\filldraw (0.25,0) circle (1.5pt) node {} ;
\filldraw (1.25,0) circle (1.5pt) node {} ;
\draw[dashed] (1.25,-0.0) arc (30:150:0.57);
\draw[dashed] (0.25,-0.0) arc (210:330:0.57);
\end{tikzpicture}\,.
\end{align}
The $\phi$ particle self-energy ($\Pi_\phi$) corresponding to this diagram can be represented by a diagram obtained by cutting one $\phi$ propagator in the above diagram, giving 
\begin{align}
\label{eq:Pi-6-legs-kept}
\begin{tikzpicture}[baseline={-0.025cm*height("$=$")}]
\draw[draw=black,thick] (-0.35,0) -- (0.25,0);
\draw[dashed] (0.75,0) circle (0.48);
\filldraw (0.25,0) circle (1.5pt) node {} ;
\filldraw (1.25,0) circle (1.5pt) node {} ;
\draw[thick] (0.25,0) -- (1.25,0);
\draw[draw=black,thick] (1.25,0) -- (1.85,0);
\end{tikzpicture} \,,
\end{align}
which we denote as $\Pi_{\phi,(6)}$.
Above, we have kept the external legs for the purpose of illustrating the cutting rules. Although the external legs do not contribute to the explicit expression of the self-energy, they represent an external $\phi$ particle in the interaction processes we are discussing for the reason that will be clear shortly. The collision term in Eq.~\eqref{eq:kinetic-eq} corresponding to $\Pi_{\phi,(6)}$ is schematically represented by (consider the loss term for example, the gain term is just the inverse process)
\begin{align}
\frac{\i}{2} \overbar{\Pi}_{\phi,(6)}^< \overbar{\Delta}_\phi^> \qquad \sim \qquad \underbrace{\begin{tikzpicture}[baseline={-0.025cm*height("$=$")}]
\draw[draw=black,thick] (-0.75,0) -- (0.25,0);
\draw[thick,dashed] (0.25,0) --  (0.78,0.75);
\draw[thick,dashed] (0.25,0) --  (0.78,-0.75);
\filldraw (0.25,0) circle (1.5pt) node {} ;
\draw[thick] (0.25,0) -- (1.25,0);
\end{tikzpicture}}_{\overbar{\Delta}^>_\phi \ \ \times \ \ \overbar{\Pi}^<_{\phi,(6)}} \quad\Leftarrow\quad \begin{tikzpicture}[baseline={-0.025cm*height("$=$")}]
\draw[draw=black,thick] (-0.35,0) -- (0.25,0);
\draw[dashed] (0.75,0) circle (0.48);
\filldraw (0.25,0) circle (1.5pt) node {} ;
\filldraw (1.25,0) circle (1.5pt) node {} ;
\draw[thick] (0.25,0) -- (1.25,0);
\draw[draw=black,thick] (1.25,0) -- (1.85,0);
\draw[draw=black,dotted] (1-0.7,0-0.7) -- (0.5+0.7,0+0.7) ;
\end{tikzpicture}\,. 
\end{align}
Note that in $\Pi_{\phi,(6)}$, there are three propagators (c.f. Eq.~\eqref{eq:Pi-phi-6}). Thus, we see three particles corresponding to it. Now we see that the external legs that we kept in Eq.~\eqref{eq:Pi-6-legs-kept} correspond to, after cutting the diagram, the $\phi$ particle from $\overbar{\Delta}_\phi^{<,>}$ in Eq.~\eqref{eq:kinetic-eq}.

The collision term actually contains more processes that allow the particles from the cut loop to reverse their role from being an outgoing particle to being an incoming particle. This is essentially due to the relations in Eqs.~\eqref{eq:relation1} and \eqref{eq:relation2}. Physically, this is due to the fact that the plasma can provide incoming particles, which is different from the case at zero temperature. In conclusion, for $\Pi_{\phi,(6)}$, we have the cutting rules:
\begin{align}
\begin{tikzpicture}[baseline={-0.025cm*height("$=$")}]
\draw[draw=black,thick] (-0.35,0) -- (0.25,0);
\draw[dashed] (0.75,0) circle (0.48);
\filldraw (0.25,0) circle (1.5pt) node {} ;
\filldraw (1.25,0) circle (1.5pt) node {} ;
\draw[thick] (0.25,0) -- (1.25,0);
\draw[draw=black,thick] (1.25,0) -- (1.85,0);
\draw[draw=black,dotted] (1-0.7,0-0.7) -- (0.5+0.7,0+0.7) ;
\end{tikzpicture} 
\qquad \Rightarrow \qquad
\underbrace{\begin{tikzpicture}[baseline={-0.025cm*height("$=$")}]
\draw[draw=black,thick] (-0.75,0) -- (0.25,0);
\draw[thick,dashed] (0.25,0) --  (0.78,0.75);
\draw[thick,dashed] (0.25,0) --  (0.78,-0.75);
\filldraw (0.25,0) circle (1.5pt) node {} ;
\draw[thick] (0.25,0) -- (1.25,0);
\end{tikzpicture} 
\,,
\begin{tikzpicture}[baseline={-0.025cm*height("$=$")}]
\draw[draw=black,thick] (-0.75,-0.75) -- (0,0);
\draw[draw=black,thick] (-0.75,0.75) -- (0,0);
\draw[dashed] (0.,0) --  (0.75,0.75);
\draw[dashed] (0.,0) --  (0.75,-0.75);
\filldraw (0,0) circle (1.5pt) node {} ;
\end{tikzpicture} 
\,,
\begin{tikzpicture}[baseline={-0.025cm*height("$=$")}]
\draw[dashed] (-0.75,-0.75) -- (0,0);
\draw[draw=black,thick] (-0.75,0.75) -- (0,0);
\draw[black,thick] (0.,0) --  (0.75,0.75);
\draw[dashed] (0.,0) --  (0.75,-0.75);
\filldraw (0,0) circle (1.5pt) node {} ;
\end{tikzpicture} 
\,,
\begin{tikzpicture}[baseline={-0.025cm*height("$=$")}]
\draw[draw=black,thick] (-0.75,0) -- (0.25,0);
\draw[dashed] (0.25,0) --  (0.25-0.55,0.75);
\draw[dashed] (0.25,0) --  (0.25-0.55,-0.75);
\filldraw (0.25,0) circle (1.5pt) node {} ;
\draw[thick] (0.25,0) -- (1.25,0);
\end{tikzpicture} 
\,,
\begin{tikzpicture}[baseline={-0.025cm*height("$=$")}]
\draw[draw=black,thick] (-0.75,0) -- (0.25,0);
\draw[dashed] (0.25,0) --  (0.25-0.55,0.75);
\draw[black,thick] (0.25,0) --  (0.25-0.55,-0.75);
\filldraw (0.25,0) circle (1.5pt) node {} ;
\draw[dashed] (0.25,0) -- (1.25,0);
\end{tikzpicture}}_{\textrm{ subject\ to\ on-shell\ conditions}} \,.
\end{align}
Above, one cuts the diagram on the left-hand side and allows the legs from the cut loops to turn right (outgoing particle) to left (ingoing particle).

\paragraph{With condensate}
Now we consider how to obtain the collision terms for the condensate-particle type vertices using the cutting rules. We consider, for example, the fourth diagram in Eq.~\eqref{eq:Gamma2} with $\Delta_\chi\rightarrow \mathbf{\Delta}_\chi$,
\begin{align}
    \begin{tikzpicture}[baseline={-0.025cm*height("$=$")}]
\draw[draw=black,thick] (-0.5,0) circle (0.15) ;
\draw[draw=black,thick] (-0.5-0.12,0+0.12) -- (-0.5+0.12,0-0.12);
\draw[draw=black,thick] (-0.5-0.12,0-0.12) -- (-0.5+0.12,0+0.12);
\draw[draw=black,thick] (-0.35,0) -- (0.25,0);
\draw[dashed] (0.75,0) circle (0.48);
\filldraw (0.27,0) circle (1.5pt) node {} ;
\filldraw (1.23,0) circle (1.5pt) node {} ;
\draw[thick] (0.25,0) -- (1.25,0);
\draw[draw=black,thick] (1.25,0) -- (1.85,0);
\draw[draw=black,thick] (2,0) circle (0.15) ;
\draw[draw=black,thick] (2-0.12,0+0.12) -- (2+0.12,0-0.12);
\draw[draw=black,thick] (2-0.12,0-0.12) -- (2+0.12,0+0.12);
\end{tikzpicture}
\qquad \Rightarrow \qquad
\text{self-energy\ } \Pi_{\phi,(4)}: \quad
\begin{tikzpicture}[baseline={-0.025cm*height("$=$")}]
\draw[draw=black,thick] (-0.35,0) -- (0.25,0);
\draw[draw=black,thick] (0.25,-0.6) circle (0.15) ;
\draw[draw=black,thick] (-0.5-0.12+0.75,0+0.12-0.6) -- (-0.5+0.12+0.75,0-0.12-0.6);
\draw[draw=black,thick] (-0.5-0.12+0.75,0-0.12-0.6) -- (-0.5+0.12+0.75,0+0.12-0.6);
\draw[draw=black,thick] (0.25,0) -- (0.25,-0.45);
\draw[dashed] (0.75,0) circle (0.5);
\filldraw (0.25,0) circle (1.5pt) node {} ;
\filldraw (1.25,0) circle (1.5pt) node {} ;
\draw[draw=black,thick] (1.25,-0.6) circle (0.15) ;
\draw[draw=black,thick] (-0.5-0.12+0.75+1,0+0.12-0.6) -- (-0.5+0.12+0.75+1,0-0.12-0.6);
\draw[draw=black,thick] (-0.5-0.12+0.75+1,0-0.12-0.6) -- (-0.5+0.12+0.75+1,0+0.12-0.6);
\draw[draw=black,thick] (1.25,0) -- (1.85,0);
\draw[draw=black,thick] (1.25,0) -- (1.25,-0.45);
\end{tikzpicture}\,.
\end{align}
Again, we have kept the external legs in the diagram for the self-energy. One then cuts the loop in the self-energy diagram. Now, we also have two possibilities for the condensate quantum: incoming condensate quantum or outgoing condensate quantum, corresponding to the absorption and emission cases discussed in section~\ref{sec:collision-with-condensate}. Note the first type discussed there, where the condensate only modifies the vertex amplitude, does not contribute if one takes the cycle average, and therefore we do not consider it. 

In conclusion, for the self-energy $\Pi_{\phi,(4)}$, we have the following cutting rules:
\begin{align}
\begin{tikzpicture}[baseline={-0.025cm*height("$=$")}]
\draw[dotted,draw=black] (0.5,-0.8) -- (1.1,0.8);
\draw[draw=black,thick] (-0.35,0) -- (0.25,0);
\draw[draw=black,thick] (0.25,-0.6) circle (0.15) ;
\draw[draw=black,thick] (-0.5-0.12+0.75,0+0.12-0.6) -- (-0.5+0.12+0.75,0-0.12-0.6);
\draw[draw=black,thick] (-0.5-0.12+0.75,0-0.12-0.6) -- (-0.5+0.12+0.75,0+0.12-0.6);
\draw[draw=black,thick] (0.25,0) -- (0.25,-0.45);
\draw[dashed] (0.75,0) circle (0.5);
\filldraw (0.25,0) circle (1.5pt) node {} ;
\filldraw (1.25,0) circle (1.5pt) node {} ;
\draw[draw=black,thick] (1.25,-0.6) circle (0.15) ;
\draw[draw=black,thick] (-0.5-0.12+0.75+1,0+0.12-0.6) -- (-0.5+0.12+0.75+1,0-0.12-0.6);
\draw[draw=black,thick] (-0.5-0.12+0.75+1,0-0.12-0.6) -- (-0.5+0.12+0.75+1,0+0.12-0.6);
\draw[draw=black,thick] (1.25,0) -- (1.85,0);
\draw[draw=black,thick] (1.25,0) -- (1.25,-0.45);
\end{tikzpicture}\,
\qquad \Rightarrow \qquad
\underbrace{
\begin{tikzpicture}[baseline={-0.025cm*height("$=$")}]
\draw[draw=black,thick] (-0.75,0) -- (0.25,0);
\filldraw (0.25,0) circle (1.5pt) node {} ;
\draw[dashed] (0.25,0) -- (1,0.75);
\draw[dashed] (0.25,0) -- (1,-0.75);
\draw[draw=black,thick] (0.25,0) -- (1.1,0);
\draw[draw=black,thick] (0.25+1,0) circle (0.15) ;
\draw[draw=black,thick] (0.13+1,-0.48+0.6) -- (0.37+1,-0.72+0.6);
\draw[draw=black,thick] (0.13+1,-0.72+0.6) -- (0.37+1,-0.48+0.6);
\end{tikzpicture}
\,,
\begin{tikzpicture}[baseline={-0.025cm*height("$=$")}]
\draw[draw=black,thick] (-0.75,0) -- (0.25,0);
\filldraw (0.25,0) circle (1.5pt) node {} ;
\draw[dashed] (0.25,0) -- (-0.5,-0.75);
\draw[dashed] (0.25,0) -- (1,-0.75);
\draw[draw=black,thick] (0.25,0) -- (1.1,0);
\draw[draw=black,thick] (0.25+1,0) circle (0.15) ;
\draw[draw=black,thick] (0.13+1,-0.48+0.6) -- (0.37+1,-0.72+0.6);
\draw[draw=black,thick] (0.13+1,-0.72+0.6) -- (0.37+1,-0.48+0.6);
\end{tikzpicture}\,,
\begin{tikzpicture}[baseline={-0.025cm*height("$=$")}]
\draw[draw=black,thick] (-0.75,0) -- (0.25,0);
\filldraw (0.25,0) circle (1.5pt) node {} ;
\draw[dashed] (0.25,0) -- (1,0.75);
\draw[dashed] (0.25,0) -- (1,-0.75);
\draw[draw=black,thick] (0.25,0) -- (-0.25,-0.5);
\draw[draw=black,thick] (0.25-0.6,-0.6) circle (0.15) ;
\draw[draw=black,thick] (0.13-0.6,-0.48) -- (0.37-0.6,-0.72);
\draw[draw=black,thick] (0.13-0.6,-0.72) -- (0.37-0.6,-0.48);
\end{tikzpicture}
\,,
\begin{tikzpicture}[baseline={-0.025cm*height("$=$")}]
\draw[draw=black,thick] (-0.75,0) -- (0.25,0);
\filldraw (0.25,0) circle (1.5pt) node {} ;
\draw[dashed] (0.25,0) -- (-0.5,0.75);
\draw[dashed] (0.25,0) -- (1.25,0);
\draw[draw=black,thick] (0.25,0) -- (-0.25,-0.5);
\draw[draw=black,thick] (0.25-0.6,-0.6) circle (0.15) ;
\draw[draw=black,thick] (0.13-0.6,-0.48) -- (0.37-0.6,-0.72);
\draw[draw=black,thick] (0.13-0.6,-0.72) -- (0.37-0.6,-0.48);
\end{tikzpicture}
}_{\text{subject\ to\ one-shell\ conditions}}\,.
\end{align}
The first/last two diagrams represent the emission/absorption of a condensate quantum by the $\phi$ particle. Note that not all the processes are possible; one has to make sure that they are kinematically allowed.

\subsection{Condensate kinetic equation}

We now consider the kinetic equation for the condensate, Eq.~\eqref{eq:Boltzman-condensate}. Again, we use the fourth diagram in Eq.~\eqref{eq:Gamma2} to illustrate the cutting rules. The condensate self-energy ($\Pi_\varphi$) corresponding to this diagram is obtained by cutting one condensate quantum. However, to illustrate the cutting rule for the condensate dissipation, we should still use the diagram in the effective action (this is similar to how we kept the external legs in drawing the diagrams for $\Pi_\phi$). Then the collision terms in Eq.~\eqref{eq:Boltzman-condensate} (corresponding to $\hat{\C}_\pi$) can be obtained from Eq.~\eqref{eq:ImPi-varphi} through the following cutting rule:
\begin{align}
\begin{tikzpicture}[baseline={-0.025cm*height("$=$")}]
\draw[draw=black,thick] (-0.5,0) circle (0.15) ;
\draw[draw=black,thick] (-0.5-0.12,0+0.12) -- (-0.5+0.12,0-0.12);
\draw[draw=black,thick] (-0.5-0.12,0-0.12) -- (-0.5+0.12,0+0.12);
\draw[draw=black,thick] (-0.35,0) -- (0.25,0);
\draw[dashed] (0.75,0) circle (0.48);
\filldraw (0.27,0) circle (1.5pt) node {} ;
\filldraw (1.23,0) circle (1.5pt) node {} ;
\draw[thick] (0.25,0) -- (1.25,0);
\draw[draw=black,thick] (1.25,0) -- (1.85,0);
\draw[draw=black,thick] (2,0) circle (0.15) ;
\draw[draw=black,thick] (2-0.12,0+0.12) -- (2+0.12,0-0.12);
\draw[draw=black,thick] (2-0.12,0-0.12) -- (2+0.12,0+0.12);
\draw[black,dotted] (0.75-0.7,0-0.7) -- (0.75+0.7,0+0.7);
\end{tikzpicture}
\qquad\Rightarrow\qquad
\underbrace{\begin{tikzpicture}[baseline={-0.025cm*height("$=$")}]
\draw[draw=black,thick] (-0.5,0) circle (0.15) ;
\draw[draw=black,thick] (-0.5-0.12,0+0.12) -- (-0.5+0.12,0-0.12);
\draw[draw=black,thick] (-0.5-0.12,0-0.12) -- (-0.5+0.12,0+0.12);
\draw[draw=black,thick] (-0.35,0) -- (0.25,0);
\filldraw (0.27,0) circle (1.5pt) node {} ;
\draw[thick] (0.25,0) -- (1.25,0);
\draw[dashed] (0.25,0) --  (0.78,0.75);
\draw[dashed] (0.25,0) --  (0.78,-0.75);
\end{tikzpicture}
\,,
\begin{tikzpicture}[baseline={-0.025cm*height("$=$")}]
\draw[draw=black,thick] (-0.5,0) circle (0.15) ;
\draw[draw=black,thick] (-0.5-0.12,0+0.12) -- (-0.5+0.12,0-0.12);
\draw[draw=black,thick] (-0.5-0.12,0-0.12) -- (-0.5+0.12,0+0.12);
\draw[draw=black,thick] (-0.35,0) -- (0.25,0);
\filldraw (0.27,0) circle (1.5pt) node {} ;
\draw[thick] (0.25,0) -- (1.25,0);
\draw[dashed] (0.25,0) --  (0.78,0.75);
\draw[dashed] (0.25-0.53,0.75) --  (0.25,0);
\end{tikzpicture}
\,,
\begin{tikzpicture}[baseline={-0.025cm*height("$=$")}]
\draw[draw=black,thick] (-0.5,0) circle (0.15) ;
\draw[draw=black,thick] (-0.5-0.12,0+0.12) -- (-0.5+0.12,0-0.12);
\draw[draw=black,thick] (-0.5-0.12,0-0.12) -- (-0.5+0.12,0+0.12);
\draw[draw=black,thick] (-0.35,0) -- (0.25,0);
\filldraw (0.27,0) circle (1.5pt) node {} ;
\draw[thick] (0.25,0) -- (1.25,0);
\draw[dashed] (0.25-0.53,-0.75) --  (0.25,0);
\draw[dashed] (0.25-0.53,0.75) --  (0.25,0);
\end{tikzpicture}
\,,
\begin{tikzpicture}[baseline={-0.025cm*height("$=$")}]
\draw[draw=black,thick] (-0.5,0) circle (0.15) ;
\draw[draw=black,thick] (-0.5-0.12,0+0.12) -- (-0.5+0.12,0-0.12);
\draw[draw=black,thick] (-0.5-0.12,0-0.12) -- (-0.5+0.12,0+0.12);
\draw[draw=black,thick] (-0.35,0) -- (0.25,0);
\filldraw (0.27,0) circle (1.5pt) node {} ;
\draw[dashed] (0.25,0) -- (1.25,0);
\draw[black,thick] (0.25-0.53,-0.75) --  (0.25,0);
\draw[dashed] (0.25-0.53,0.75) --  (0.25,0);
\end{tikzpicture}}_{\textrm{subject\ to\ on-shell\ conditions}}\,.
\end{align}
For $\hat{\C}_v$, one simply considers the following diagram
\begin{align}
    \begin{tikzpicture}[baseline={0cm-0.5*height("$=$")}]
\draw[thick, dashed] (0,0) circle (0.5) ;
\filldraw (0.5,0) circle (1.5pt) node {} ;
\filldraw (-0.5,0) circle (1.5pt) node {} ;
\draw[thick] (0.5,0) -- (0.6+0.3*0.707,0.1+0.3*0.707) ;
\draw[thick] (0.6+0.45*0.707,0.1+0.45*0.707) circle (0.15) ;
\draw[thick] (0.6+0.3*0.707,0.1+0.3*0.707) -- (0.6+0.6*0.707,0.1+0.6*0.707) ;
\draw[thick] (0.6+0.6*0.707,0.1+0.3*0.707) -- (0.6+0.3*0.707,0.1+0.6*0.707) ;
\draw[thick] (0.5,0) -- (0.6+0.3*0.707,-0.1-0.3*0.707) ;
\draw[thick] (0.6+0.45*0.707,-0.1-0.45*0.707) circle (0.15) ;
\draw[thick] (0.6+0.3*0.707,-0.1-0.3*0.707) -- (0.6+0.6*0.707,-0.1-0.6*0.707) ;
\draw[thick] (0.6+0.6*0.707,-0.1-0.3*0.707) -- (0.6+0.3*0.707,-0.1-0.6*0.707) ;
\draw[thick] (-0.5,0) -- (-0.6-0.3*0.707,-0.1-0.3*0.707) ;
\draw[thick] (-0.6-0.45*0.707,-0.1-0.45*0.707) circle (0.15) ;
\draw[thick] (-0.6-0.3*0.707,-0.1-0.3*0.707) -- (-0.6-0.6*0.707,-0.1-0.6*0.707) ;
\draw[thick] (-0.6-0.6*0.707,-0.1-0.3*0.707) -- (-0.6-0.3*0.707,-0.1-0.6*0.707) ;
\draw[thick] (-0.5,0) -- (-0.6-0.3*0.707,0.1+0.3*0.707) ;
\draw[thick] (-0.6-0.45*0.707,0.1+0.45*0.707) circle (0.15) ;
\draw[thick] (-0.6-0.3*0.707,0.1+0.3*0.707) -- (-0.6-0.6*0.707,0.1+0.6*0.707) ;
\draw[thick] (-0.6-0.6*0.707,0.1+0.3*0.707) -- (-0.6-0.3*0.707,0.1+0.6*0.707) ;
\end{tikzpicture}\,,
\end{align}
which is obtained from the $\chi$ one-loop diagram after the small-field expansion in $\varphi$.

\bibliographystyle{utphys}
\bibliography{ref}

\end{document}